\documentclass{aa}
\usepackage{graphicx}
\usepackage{natbib}
\bibpunct{(}{)}{;}{a}{}{,}
\usepackage{txfonts}
\begin{document}

   \title{The clustering of merging star-forming haloes: dust emission as high frequency arcminute CMB foreground}
   \titlerunning{The clustering of merging star-forming haloes: CMB foreground}
   \authorrunning{M. Righi et al.}

   \author{M. Righi
          \inst{1},
          C. Hern\'andez-Monteagudo
          \inst{2},
          \and
          R.A. Sunyaev\inst{1,3}
          }

   \offprints{M. Righi,\\ e-mail: righi@mpa-garching.mpg.de}

   \institute{Max-Planck-Institut f\"ur Astrophysik, 
              Karl-Schwarzschild-Str. 1, 85748 Garching, Germany
\and
              Department of Physics and Astronomy, University of Pennsylvania,
              209 South 33rd St, Philadelphia, PA 19104
\and
              Space Research Institute (IKI), 
              Profsoyuznaya 84/32, Moscow, 117810, Russia
             }

   \date{Received ...; ...}

  \abstract
   {Future observations of CMB anisotropies will be able to probe high multipole regions of the angular power spectrum, corresponding to a resolution of a few arcminutes. Dust emission from merging haloes is one of the foregrounds that will affect such very small scales.}
   {We estimate the contribution to CMB angular fluctuations from objects that are bright in the sub-millimeter band due to intense star formation bursts following merging episodes.}
   {We base our approach on the Lacey-Cole merger model and on the Kennicutt relation which connects the star formation rate in galaxies with their infrared luminosity. We set the free parameters of the model in order to not exceed the SCUBA source counts, the Madau plot of star formation rate in the universe and COBE/FIRAS data on the intensity of the sub-millimeter cosmic background radiation.}
   {We show that the angular power spectrum arising from the distribution of such star-forming haloes will be one of the most significant foregrounds in the high frequency channels of future CMB experiments, such as PLANCK, ACT and SPT. The correlation term, due to the clustering of multiple haloes at redshift $z\sim2-6$, is dominant in the broad range of angular scales $200\la l\la3000$. Poisson fluctuations due to bright sub-millimeter sources are more important at higher $l$, but since they are generated from the bright sources, such contribution could be strongly reduced if bright sources are excised from the sky maps. The contribution of the correlation term to the angular power spectrum depends strongly on the redshift evolution of the escape fraction of UV photons and the resulting temperature of the dust. The measurement of this signal will therefore give important information about the sub-millimeter emission and the escape fraction of UV photons from galaxies, in the early stage of their evolution.}
   {}

   \keywords{Cosmology: cosmic microwave background -- 
             Cosmology: theory -- 
             Galaxies: intergalactic medium -- 
             Infrared: galaxies}

   \maketitle

\section{Introduction}\label{sec:intro}
Three very important CMB experiments are due to operate in the next two years: the PLANCK surveyor spacecraft \citep{planck} and two ground based experiments, ACT \citep{act} and SPT \citep{spt}. They will cover the $[150-350]$~GHz frequency range with a resolution of a few arcminutes ($l=10^3-10^4$) and with unprecedented sensitivity. A lot of work has been invested in the last years in the study of possible foreground sources for these relatively high CMB frequencies. 

The galactic dust radiation is obviously the most well known foreground source \citep{schlegel,tegmark} on the large and intermediate angular scales relevant for CMB observations, but it is very important to also look for foreground sources that have significant power at small angular scales ($l\sim10^3-10^4$). Several well studied foregrounds, like clusters of galaxies, observable with the thermal and kinetic SZ effects \citep{kSZ,tSZ,kk99,chm06}, SCUBA galaxies \citep{scottwhite} and bright infrared galaxies \citep{LIRG,cib} are predicted to contribute significantly to the $C_l$ measured by the instruments mentioned above. It is relevant that these foregrounds are not only considered as difficulties impeding the study of primordial angular fluctuations, but that they might also provide us with extremely useful information on the statistical properties of objects like clusters of galaxies and compact sub-millimeter sources, and about the history of enrichment by heavy elements \citep{basu}. There are several projects, including APEX \citep{apex}, ASTE \citep{aste} and CCAT \citep{ccat}, which also plan to observe the sky in the sub-millimeter band.

In this work, we will study the consequences of a simple model in which haloes become bright in sub-millimeter wavelengths due to intense star formation resulting from mergers and the corresponding supply of fresh gas. We use the \citet{lc93} approach, based on the extended Press-Schechter formulation, to obtain the space density of the haloes in the universe. Knowing the space density of merging objects as a function of final mass and of the progenitor mass, we then introduce the lifetimes of the bright star formation phase as derived from numerical simulations of merging gas-rich disk galaxies. Using the Kennicutt relation between star formation rate and infrared luminosity \citep{kennicutt} and taking into account that the bulk of infrared luminosity of the galaxies with intense star formation peaks in the sub-millimeter band at $\lambda\sim100$~$\mu$m, we obtain the luminosity function of the merging sub-millimeter objects. The emission spectrum of optically thin dust is a strongly increasing function of frequency \citep[close to $\nu^{2+\beta}$, with $\beta=1.5$,][]{hildebrand}. Therefore distant objects are observed at a rest-frame frequency closer to the peak of their spectrum. At frequencies below 1000~GHz, there is strong K-correction and the flux from galaxies at $z>1$ remains approximately constant with increasing redshift \citep{blainrev}.

Recent observations of quasars at very high redshift ($z\sim6$) and of extremely massive galaxies at $z>4$ show that the chemical abundance of the gas (including dust) in the most dense parts of star-forming galaxies is close to or exceeds the solar value \citep{fan}. Therefore it is natural to assume that the dust radiation spectrum does not change dramatically with the redshift of the merging objects. This is the weakest assumption of the paper, which might lead to some changes in the dust emission spectrum at very high redshift when the abundance of heavy elements in gas and dust will drop below $10^{-2}-10^{-3}$ of the solar value. Nevertheless, observations of the Magellanic Clouds indicate that absorption and emission of dust remains similar to that in the Milky Way, when the abundance of chemical elements is changing by one order of magnitude \citep{ld02}.

Our simple approach permits us to obtain several predictions that can be compared with the experimental results: 
\begin{enumerate}
\item the star formation rate density in the universe as function of redshift \citep[i.e. the \emph{Madau plot},][]{madau96,madau98,hopkins04,hopkins06};
\item the upper limit on the intensity of the infrared background set by COBE/FIRAS observations \citep{fixsen98};
\item the source counts of sub-millimeter sources performed by SCUBA \citep{scuba2};
\item the redshift distributions of the brightest SCUBA sources \citep{chapman}.
\end{enumerate}

This model permits us to explain all but two of the above experiments. It demonstrates a deficiency of the brightest sources in comparison with the SCUBA data, but much more importantly it shows that the redshift distribution of the brightest objects is also significantly different. We are then forced to make the simplest conclusion that SCUBA is observing objects of a different nature. In this paper we are looking for the answer to the following question: \emph{where is the family of the sub-millimeter sources associated with galaxy mergings and why it is not detected yet in the direct source counts?} Forthcoming SPT and ACT deep surveys will be able to find an additional population of sub-millimeter sources, since at the sensitivity levels of these instruments, up to $\sim50\%$ of all detected objects should be merging galaxies. This population should have a different redshift distribution than the SCUBA objects.
Much deeper surveys with ALMA could also be dominated by sub-millimeter objects connected with merging galaxies, depending on the properties of SCUBA sources, whose exact physical nature is still a matter of debate. Ongoing observations with SPITZER and the future results of Herschel will give additional clues about the merging population in the infrared observations.

The merging model is able to produce a rather good fit to the SCUBA source counts if the dust temperature is close to 25 K, but this value is not in agreement with the numbers presented in the SCUBA papers. Additional energy coming from relatively weak AGN inside the merging objects could increase the flux from individual sources and mimic the source counts curve of SCUBA.

The results of our computations show that the dust emission of the merging objects should be one of the key foreground sources for primordial fluctuations of CMB, in the very broad range of angular scales $l\simeq10^2-10^4$. Our model demonstrates the importance of the correlation term in providing the fluctuations due to haloes at high redshift ($z\simeq2-6$) and that its contribution exceeds the predictions for the Poisson fluctuations in the number of star forming haloes. In this way we confirm earlier conclusions of \citet{haikno}, about the importance of the correlation signal, which were based on a different model of sub-millimeter sources and on a different statistical characterization of the population (number of haloes instead of number of mergers).

The correlation term from the relatively small mass ($10^8-10^{11}\,M_{\odot}$) but numerous  merging objects gives a contribution that samples the typical comoving correlation length of these objects ($\la10\,h^{-1}$Mpc). This signal probes the most overdense regions in the universe at high redshift, that eventually will give rise to clusters and superclusters of galaxies. The Poisson term is smaller than the correlation term up to $l\sim3000$, but becomes more important at higher $l$. ACT and SPT will be so sensitive that they will be able to perform successful source counts in their high frequency spectral bands which will permit the removal of the brightest sources (including mainly SCUBA sources) from the resulting maps at lower frequency, decreasing the contribution of the Poisson noise to the power spectrum of the background. This procedure will be unable to affect significantly the correlation term, which arises mainly from the numerous low-flux sources.

We model here the sub-millimeter emission arising after violent starburst episodes, with large populations of young stars appearing on very short timescales (of the order of $10^7-10^8$ years).
The level of the predicted correlated signal depends strongly on the assumptions about:
\begin{enumerate}
\item the escape fraction of UV radiation from merging star-forming galaxies and its evolution with redshift;
\item the properties of the dust emitting in the sub-millimeter band (its effective temperature at high redshift and the emissivity index $\beta$);
\item the redshift dependence of heavy elements and dust abundance;
\item the amount of gas inside merging haloes.
\end{enumerate}
All these parameters are interconnected. Observations will give some hints about them and we will discuss below the information that could be obtained by precise observations of the angular power spectrum at different frequencies.

Throughout the paper we use the recent WMAP3 cosmological parameters for standard $\Lambda$CDM cosmology \citep{wmap3}.

\section{Halo distribution model}\label{sec:halo}

To characterize the halo population we will use the standard Press-Schechter approach \citep{ps74} in the extended formulation of \citet{lc93}, which also includes a prescription for the halo mergers.

\subsection{Star formation from halo mergers}\label{subsec:sfr}
We model the star formation within the framework of the halo model and derive the star formation rate as the rate of the baryonic mass that is accreted per unit time into new haloes. 

According to \citet{lc93}, the merger rate of a halo of mass $M_1$ with another of mass $M_2$ in a final halo of mass $M=M_1+M_2$ is given by
\begin{eqnarray}\label{LC}
\frac{dN_{\mathrm{merg}}}{dMdt}(M_1\rightarrow M,t)&=&\sqrt{\frac{2}{\pi}}
\frac{1}{t}\left|\frac{d\ln\delta_{\mathrm{cr}}}{d\ln t}\right|\left|\frac{d\sigma}{dM}\right|\frac{\delta_{\mathrm{cr}}}{\sigma(M)^2}\nonumber\\
&&\times\,\frac{1}{[1-\sigma(M)^2/\sigma(M_1)^2]^{3/2}} \nonumber \\
 \nonumber \\
&&\times\,\exp\left[-\frac{\delta_{\mathrm{cr}}^2}{2}\left(\frac{1}{\sigma(M)^2}-\frac{1}{\sigma(M_1)^2}\right)\right],\nonumber\\
\end{eqnarray}
where $\delta_{\mathrm{cr}}$ is the critical density for spherical collapse and $\sigma(M)^2$ is the variance of the linear density field.\\
As pointed out by \citet{benson}, the merger model is consistent only if the ratio between the masses of the merging haloes is not larger than $\sim10-100$. Therefore in the following analysis we will include only \emph{major mergers}, i.e. such that $0.1\la M_1/M_2\la10$.

In each merging episode a given amount of gas is converted into stars. We must take into account the cooling time which, for a given temperature $T$ and metallicity $Z$, can be written as
\begin{equation}\label{tcool}
t_{\mathrm{cool}}(T,Z)=\frac{3}{2}\frac{\rho_{\mathrm{gas}}\,kT}{\mu\,\Lambda(T,Z)\,n_{\mathrm{H}}^2},
\end{equation}
where $\Lambda(T,Z)$ is the cooling function (in units of erg cm$^3$ s$^{-1}$) and $\mu$ is the mean molecular weight of the gas. The temperature of the gas is assumed to be equal to the virial temperature. The cooling functions are tabulated for different values of $T$ and $Z$ by \citet{sutherland}. For our estimate of the cooling time we choose a value $Z=Z_{\odot}$: observations of quasars and galaxies at $z\sim6$ by \citet{fan} show that the metal content of such objects is close to the solar value, therefore this approximation is valid at least to this redshift. For even more distant objects the value of the metallicity is expected to be lower, but we will see that such objects will not contribute significantly to the effects that we are aiming to model. The cooling time sets a limit for the star formation, since only haloes with $t_{\mathrm{cool}}\le t_{\mathrm{Hub}}$ can cool within the cosmological time. This excludes the very massive haloes and the haloes with a mass corresponding to a virial temperature lower than $T_{\mathrm{vir}}\simeq10^4$ K.

\begin{figure}
  \resizebox{\hsize}{!}{\includegraphics{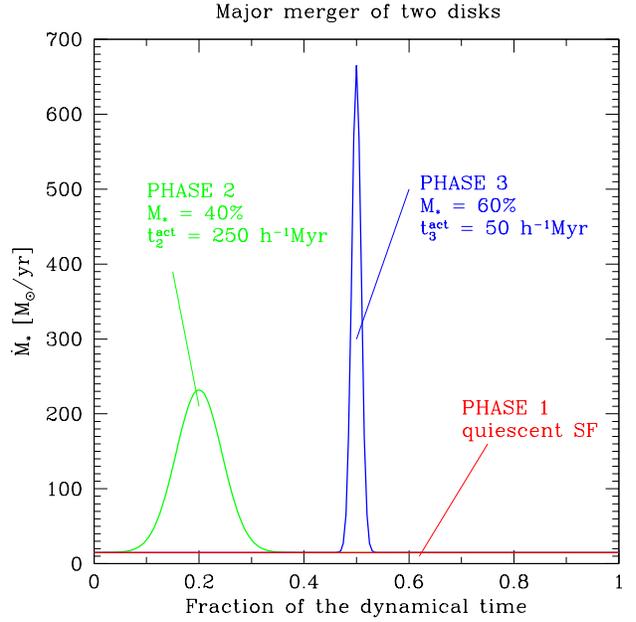}}
  \caption{The star formation rate as a function of time during a merger of two gas-rich massive disk galaxies of equal mass ($M\simeq10^{12}\,h^{-1}M_{\odot}$) used for the computations in this paper and based on the results of numerical simulations, cited in the text. Different colours identify different phases of star formation activity.}
  \label{fig:MODEL_phases}
\end{figure}

Let us now consider two haloes of mass $M_1$ and $M_2$, merging to yield a halo of mass $M=M_1+M_2$. In each episode a given amount of the halo mass is converted into stellar mass.
We parametrize this in the following way
\begin{equation}
M_{\star}^1=\frac{\Omega_{\mathrm{b}}}{\Omega_{\mathrm{m}}}\,\eta\,M_1\frac{M_2}{M/2}.
\end{equation}
The amount of stellar mass formed from the halo $M_1$ is therefore proportional to its mass and to the mass of the merging halo $M_2$, if this is sufficiently massive. In the same way, for the second halo
\begin{equation}
M_{\star}^2=\frac{\Omega_{\mathrm{b}}}{\Omega_{\mathrm{m}}}\,\eta\,M_2\frac{M_1}{M/2}.
\end{equation}
The total stellar mass produced in this merging episode is then the sum
\begin{equation}
M_{\star}=M_{\star}^1+M_{\star}^2=4\,\frac{\Omega_{\mathrm{b}}}{\Omega_{\mathrm{m}}}\,\eta\,\frac{M_1\cdot M_2}{M}.
\end{equation}
The parameter $\eta$ is the star formation efficiency. We take this to be $5\%$, in order to match the observations of the cosmic star formation history, which will be presented in Section~\ref{subsec:madau}.\\
Using this simple parametrization we are able to derive the stellar mass produced in a merger process between two haloes of given mass.

\subsection{Lifetime of the starburst phase}
The approach described above allows us to compute the stellar mass produced in each merger episode. Since we are interested in deriving the rate at which gas is converted into stars and then connecting it to the luminosity of the sources, we introduce a characteristic timescale for star formation.

Several numerical simulations of merging disk galaxies \citep{MH94,MH96,SH05,ROB06} show that the star formation process that occurs during a merger consists of multiple starburst episodes. Tidal interactions between merging galaxies can trigger star formation in the two disks before the final coalescence, although this usually coincides with the strongest burst. An accurate analytical description of such a process is complicated and thus numerical techniques must be used to find a solution. In the following, we will introduce a phenomenological prescription for the star formation in the merging haloes, based on the results of the numerical simulations mentioned above. These works studied the merging of two equal mass gas-rich galaxies and predicted an evolution of the star formation rate with time characterized by several phases, as displayed in Figure~\ref{fig:MODEL_phases}:

\begin{enumerate}
\item a long underlying phase (red), with a low and constant star formation rate, which is independent of the major merging process and is the sum of the quiescent star formation rates in each disk prior to the encounter;

\item a second, more active phase (green), which corresponds to the first close passage of one galaxy around the other, when the tidal deformation experienced by each disk creates shocks and compression of the gas. This triggers a moderate burst of activity of about $200\,M_{\odot}/\mathrm{yr}$, with a duration of the order of $3\times10^8$ years;

\item a strong starburst phase (blue), which occurs when the two galaxies finally merge, with a star formation rate higher than $600\,M_{\odot}/\mbox{yr}$ over a very short timescale, lower than $10^8$~years.
\end{enumerate}
We will take this model to be representative of the star formation process during mergers of massive gas-rich objects. The bulk of the objects that contribute to the merging activity are at rather high redshift ($z\ga0.3-0.5$), where the fraction of spirals and irregulars is higher than in our vicinity, where ellipticals are already a significant part of the total.

As we have seen in Section~\ref{subsec:sfr}, each merger between two haloes $M_1$ and $M_2$ produces a certain amount of stellar mass $M_{\star}(M_1,M,z)$. Such stellar mass will be created in the two active phases identified in Figure~\ref{fig:MODEL_phases} in green and blue, according to the parameters summarized in the same plot. The first phase does not involve the stellar mass produced by the merger and hence will not be included in the model. Nevertheless, in Section~\ref{subsec:spir} we will estimate the contribution of this stage to the correlation term of the angular power spectrum.

We therefore have $M_{\star}=M_{\star}^{(2)}+M_{\star}^{(3)}$, with $M_{\star}^{(2)}=0.40\,M_{\star}$ and $M_{\star}^{(3)}=0.60\,M_{\star}$. The percentages of stellar mass formed in each event are based on an average of the results presented in the literature and mentioned at the beginning of this section.\\
Introducing the active time of each phase, $t^{(i)}_{\mathrm{act}}$, it is easy to derive the corresponding star formation rate
\begin{equation}\label{mdot}
\dot{M}_{\star}^{(i)}=\frac{M_{\star}^{(i)}}{t^{(i)}_{\mathrm{act}}},\qquad i=2,3.
\end{equation}
where $t^{(2)}_{\mathrm{act}}=2.5\times10^8\,h^{-1}$yr and $t^{(3)}_{\mathrm{act}}=5\times10^7\,h^{-1}$yr.

\begin{figure}
  \resizebox{\hsize}{!}{\includegraphics{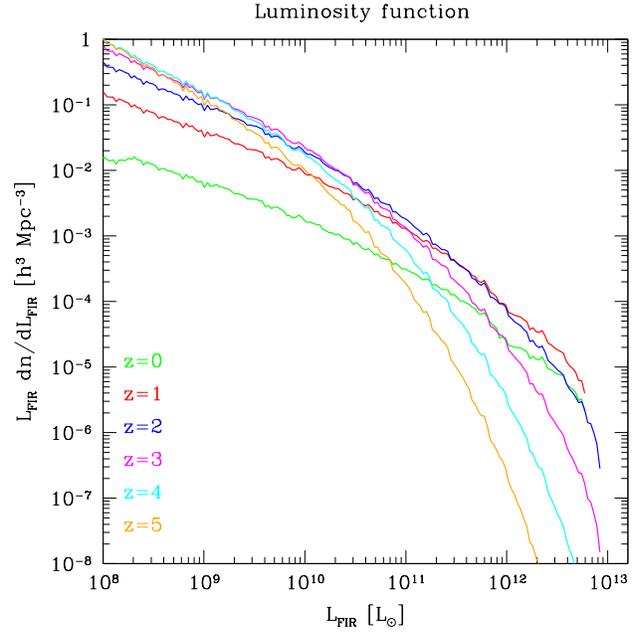}}
  \caption{The shape of the far-infrared luminosity function of the merging haloes at different redshifts.}
  \label{fig:lumfunc}
\end{figure}

\begin{figure}
  \resizebox{\hsize}{!}{\includegraphics{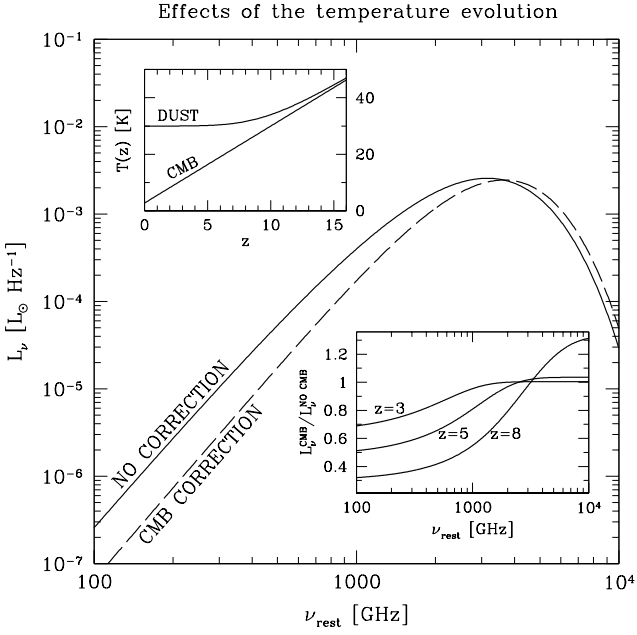}}
\caption{The effect of the correction in Equation~(\ref{dustspec}) on the spectrum of star forming haloes at $z=10$: the solid line is the spectrum with dust greybody only, the dashed line is the spectrum when including the correction due to the CMB.  In the bottom insert the ratio between the two spectra at differrent redshifts is displayed to stress the importance of the correction with increasing redshift. The top insert shows the dust and CMB temperature evolution according to Equation~(\ref{tempz}).}
  \label{fig:speccorr}
\end{figure}

\subsection{Infrared luminosity of the haloes}
The star formation rate given in (\ref{mdot}) is strongly correlated with the infrared luminosity of the source, through the Kennicutt relation \citep{kennicutt}
\begin{equation}\label{kennic}
\dot{M}_{\star}\,[M_{\odot}/\mathrm{yr}]=1.71\times10^{-10}\,L_{\mathrm{IR}}\,[L_{\odot}],
\end{equation}
where IR indicates the $[8-1000]$~$\mu$m band. The bulk of emission in this band peaks at about $100$~$\mu$m  and around $80\%$ of it is due to low temperature dust \citep{cib}. Therefore in our estimate we can directly use this relation and apply it to obtain the luminosity due to low temperature dust in star-forming galaxies.\\
The number density of such bright objects for each population can be derived with the aforementioned model as
\begin{equation}\label{nbright}
\frac{dN_{\mathrm{bright}}^{(i)}}{dV}(\dot{M}_{\star},z)=\frac{dN_{\mathrm{merg}}}{dVdt}(M,z)\cdot t_{\mathrm{act}}^{(i)},
\end{equation}
where $\frac{dN_{\mathrm{merg}}}{dVdt}(M_1,M,z)$ is the number density of mergers per unit time, which can be obtained weighting Equation~(\ref{LC}) over the mass function of the merging haloes
\begin{equation}
\frac{dN_{\mathrm{merg}}}{dVdt}(M,z)=\int dM_1\int dM \frac{dn}{dM_1}\frac{dN_\mathrm{merg}}{dMdt},
\end{equation}
remembering that $M=M_1+M_2$ and that $\dot{M_{\star}}$ depends on both the masses of the merging haloes.\\
The total number of bright objects at a given redshift will then be obtained by summing over the populations
\begin{equation}\label{nbright3}
\frac{dN_{\mathrm{bright}}}{dV}(\dot{M}_{\star},z)=\frac{dN_{\mathrm{bright}}^{(2)}}{dV}(\dot{M}_{\star},z)+\frac{dN_{\mathrm{bright}}^{(3)}}{dV}(\dot{M}_{\star},z).
\end{equation}
Equation~(\ref{nbright3}) is a differential number density of bright objects in the universe. This is a crucial quantity, especially if we replace $\dot{M}_{\star}$ with the infrared luminosity by taking advantage of the Kennicutt relation. In this way we obtain a statistical determination of the sources responsible for dust emission, i.e. a luminosity function for these objects, which is shown in Figure~\ref{fig:lumfunc}. The sharp cutoff at $L_{\mathrm{IR}}\simeq10^{13}\,L_{\odot}$ is due to the cooling time, which prevents the formation of the most massive (most luminous) objects.

\begin{figure*}
\centering
  \includegraphics[width=8.2cm]{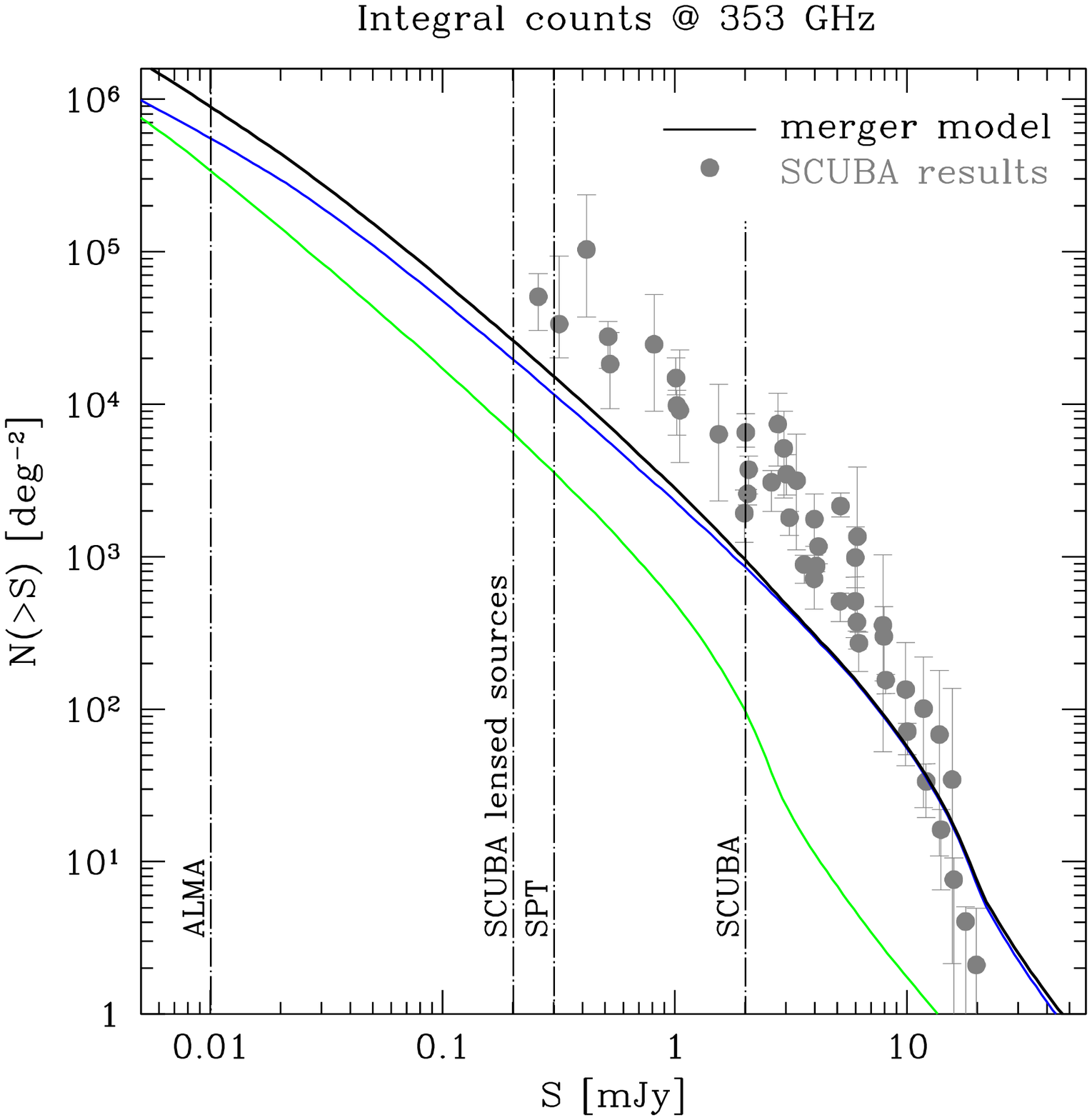}
  \includegraphics[width=8.2cm]{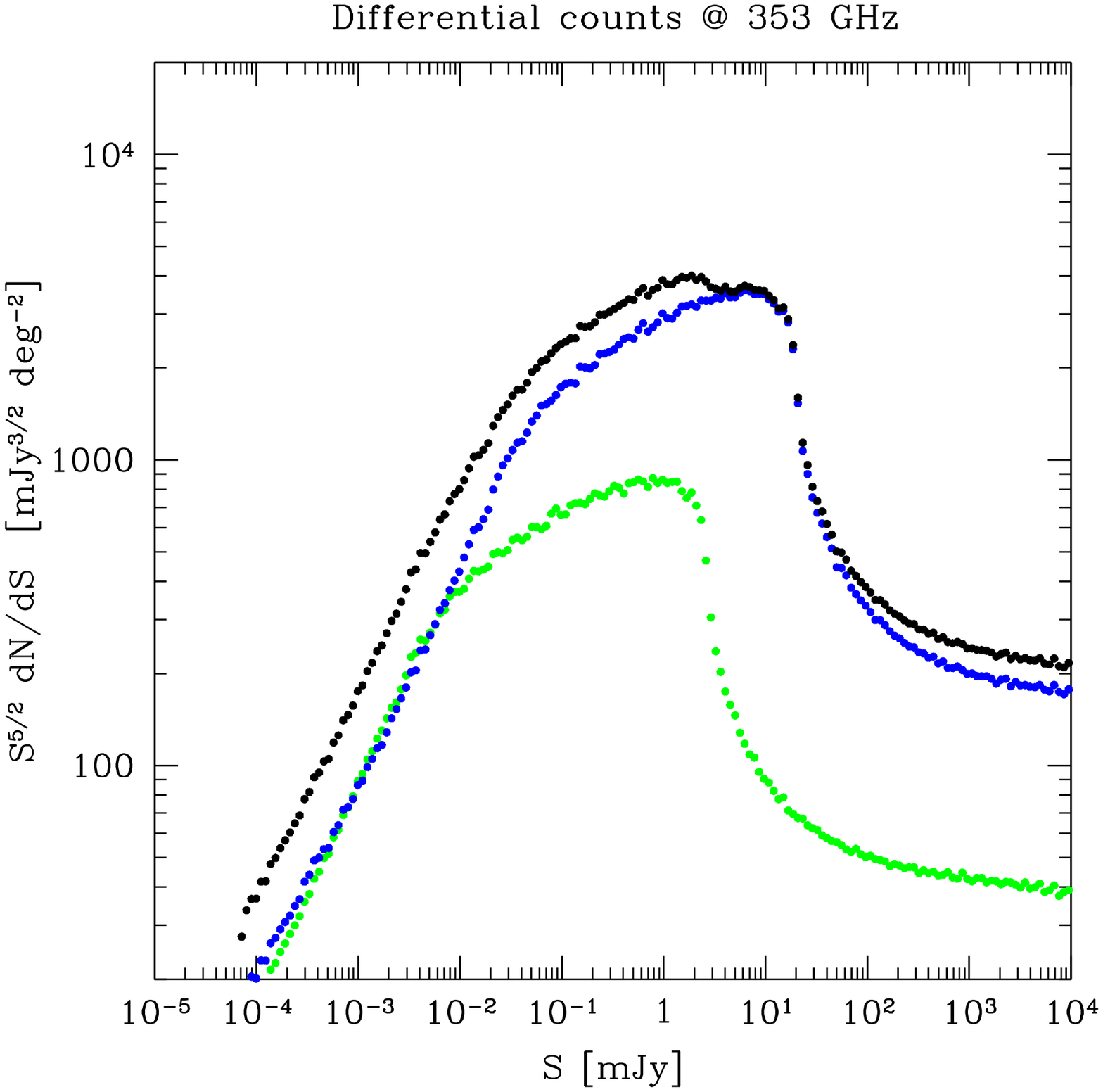}
  \caption{Testing our model with the SCUBA results. Different colours
  refer to the populations generated by the two bursts of star
  formation, as shown in Figure~\ref{fig:MODEL_phases} (green and
  blue). Both plots assume $T_{\mathrm{dust}}=30$ K (with the CMB
  correction discussed in Section~\ref{subsec:speccorr}), $\beta=1.5$
  and a mass-dependent escape fraction, which will be discussed in Section~\ref{subsubsec:fesc}. \emph{Left}: the integral counts predicted by our model (black solid line) at 353~GHz (850~$\mu$m), compared with the compilation of \citet{scuba2} for SCUBA sources (dots). The vertical lines represents the sensitivity limits of SCUBA \citep{scuba2,knudsen}, SPT and ALMA \citep[expected in $\sim$ 3 hours integration;][]{spt,alma}. \emph{Right}: the differential counts at 353~GHz. We see that the short active phase is the most important and dominates the source counts at the high-flux limit. At the low fluxes the contribution from the two bursts is pratically equal.}
  \label{fig:scubatest}
\end{figure*}

\subsection{Angular fluctuations from the star-forming objects}
The angular distribution of the emission of these star-forming sources will be characterised by their spatial distribution at different cosmic epochs. These sources will be located in the highest density regions in the universe, and their spatial clustering properties are well determined for a given cosmological scenario. Since most of the contribution comes from sources located at high redshift, their clustering properties turn out to be particularly relevant at angular scales of a few arcminutes. The shot noise associated with the Poissonian character of the source distribution will be relevant only at smaller angular scales.\\
In Appendix~\ref{app:powerspec} we provide a detailed computation of the statistical properties of the angular intensity fluctuations generated by the population of star-forming objects at high redshift. We find that, when probing sufficiently large scales, the change of brightness can be written, to linear order in the perturbations, as an integral along the line of sight
\begin{eqnarray}\label{deltaexp}
\Delta I_{\nu} = \mbox{const}&+&\int_{0}^{r_{\mathrm{LSS}}}dr \left[\frac{\vec{v}\cdot\vec{n}}{c}\,a(r)\,\bar{n}(r)\,\frac{L_{\nu(1+z[r])}}{4\pi}+2\,\frac{\dot\phi}{c}\,\bar{I}_{\nu}^{\mathrm{dust}}(r)\right]\nonumber \\
&+&\int_{0}^{r_{\mathrm{LSS}}} dr \,a(r)\,\delta n(\vec{r})\,\frac{L_{\nu(1+z[r])}}{4\pi}\,a^3(r)+\mathcal{O}(\delta^2),\nonumber\\
\end{eqnarray}
where $r$ is the comoving distance, $a(r)$ the scale factor and $n(r)$ the number density of sources. The first term in the rhs is constant and introduces no anisotropy. The first integral contains the Doppler contribution, due to the peculiar velocity $\vec{v}$ of the sources, and the Integrated Sachs Wolfe (ISW) blueshift, due to time-varying potential $\dot\phi$, that the emitted photons experience as they travel towards the observer. These two contributions are found to be negligible when compared to the second integral, which provides the contribution introduced by the varying spatial distribution of sources.  This term can be separated in the contribution from the so-called {\it Poisson} term, accounting for the Poissonian nature of the objects, and from the {\it correlation} term, accounting for dependence of the source number density on the environment density.

\section{A model for dust emission}\label{sec:dust}
A significant fraction of the stellar radiation emitted by galaxies is reprocessed by dust and reradiated in the infrared and sub-millimeter band. The dust in star-forming galaxies has a very complicated distribution in space and a broad range of temperatures, depending on the proximity of dust grains to the sources of ionizing, optical and short-wavelength infrared radiation. The emissivity of the dust depends strongly on its composition and on the size distribution of grains. We do not have enough observational data to construct a reliable detailed model connecting the sub-millimeter spectra of the galaxies at redshift $z>2$ with the star formation inside them. Several observations \citep{blaint,blainrev,chapman} show that in the majority of the brightest extragalactic infrared objects and in particular in SCUBA galaxies, the observed spectrum in the [100-3000]~$\mu$m band is well fitted by a single-temperature modified blackbody $\nu^{\beta}B_{\nu}(T_{\mathrm{dust}})$ with a temperature $T_{\mathrm{dust}}\simeq30-50$~K and an emissivity index in the range $1\le\beta\le2$; there is little change in the properties of dust along the Hubble sequence, thus we can expect this fit to be valid independently of the Hubble type. As it is obvious from the discussions in the papers of \citet{reach} and \citet{schlegel}, a fit with $\beta$ significantly smaller than 2 is attributable to multi-temperature dust or a difference in chemical composition.

For galaxies at high-redshift having similar rates of dust heating, the increase in the temperature of the cosmic microwave background will affect the dust temperature. Therefore if the dust temperature in an object at $z=0$ is $T_{\mathrm{dust}}$, for a similar object at redshift $z$ it would increase as \citep{blaint}
\begin{equation}\label{tempz}
T_{\mathrm{dust}}(z)=\left[T_{\mathrm{dust}}^{4+\beta}+T_{\mathrm{cmb}}^{4+\beta}\left((1+z)^{4+\beta}-1\right)\right]^{1/(4+\beta)},
\end{equation}
where $T_{\mathrm{cmb}}=2.725$ K is the temperature of the background today.

We consider the observations by PLANCK, SPT and ACT at fixed frequencies, namely 147, 217, 274 and 353~GHz; the peak of the SED for the majority of the brightest SCUBA and infrared galaxies is around 3000~GHz. This means that the frequencies higher than the peak will begin to dominate these observing channels at redshifts $z\ga19$, 13, 10 and 8, respectively, so the flux observed in a given band will correspond to the flux emitted at a frequency higher than the peak. This has two consequencies. First, the contribution from high redshift objects will begin to decrease in the highest frequency bands of PLANCK, ACT and SPT. But in moderate frequency channels (147 and 217~GHz) these experiments continue to be sensitive to the first merging galaxies in the universe. Furthermore, for the high frequency channels the approximation of hot single-temperature dust emission will not be accurate, since in observations of the local bright objects the region of wavelengths shorter than 100~$\mu$m shows that the decline in flux has a power-law behaviour, rather than the exponential cutoff of the single-temperature model. This demonstrates that the dust in reality has a broad distribution of temperatures which makes the cutoff after the peak much less steep.

\subsection{Dust emission in the CMB thermal bath at high redshift}\label{subsec:speccorr}
The fit by a modified blackbody spectrum considered above is valid under the further assumption that the influence of the CMB radiation field on dust grains can be neglected. Such an approximation is good at low redshift, when the background temperature is low, but as the redshift increases the intensity of this radiation becomes non-negligible. Therefore it is necessary to introduce a correction in the assumed spectral energy distribution of infrared sources
\begin{equation}\label{dustspec}
L_{\nu}\propto\nu^{\beta}\left[B_{\nu}(T_{\mathrm{dust}})-B_{\nu}(T_{\mathrm{cmb}})\right],
\end{equation}
where the difference in square brackets represents the correction due to the presence of the CMB thermal bath. In the low frequency limit ($h\nu\ll kT_{\mathrm{dust}}$) the luminosity is proportional to the difference between the grain temperature and the temperature of the CMB. As a result, for the same optical and ultraviolet energy absorbed by the grain, its radiation in the high frequency band will increase due to the existence of the CMB thermal bath and correspondingly will decrease in the low frequency part of the spectrum. This is plotted in Figure~\ref{fig:speccorr} for an object at $z=10$ with infrared luminosity $L_{\mathrm{IR}}=10^{10}L_{\odot}$.\\

\section{Three observational tests}\label{sec:test}
We identified the merger formalism presented in Section~\ref{sec:halo} as the main process leading to star formation in galaxies. Within the framework provided by the standard $\Lambda$CMD Cosmology, this approach constitutes a model for the IR emission in the Universe that can be compared with three different observational constraints, as we show below.

The model is particularly sensitive to the two spectral parameters, the emissivity index $\beta$ and the temperature $T_{\mathrm{dust}}$ of dust, which convert the bolometric luminosity into the spectral luminosity and hence fix the distribution of the infrared sources. Another important quantity, which will be discussed in this section, is the escape fraction of ionizing photons, which determines the fraction of ultraviolet radiation absorbed by dust and hence sets the relation between the star formation rate and the infrared luminosity. We will see that the introduction of such a parameter is necessary to satisfy the limits on the intensity of the infrared background set by the COBE/FIRAS data. Several observational tests can be performed to calibrate the free parameters of the model.

\subsection{SCUBA source counts}
Our merging halo model provides the source counts for the sub-millimeter sources. Comparison of our counts with SCUBA observations \citep{scuba2} at 850~$\mu$m (353~GHz) shows that, in the region above the SCUBA sensitivity limit at 0.2 mJy, the curve is rather sensitive to the dust emission spectral parameters. In the left panel of Figure~\ref{fig:scubatest} we plot the source counts resulting from our model. We will show in Section~\ref{subsec:sensparam} the variation of the curves for different values of the temperature and emissivity index: the closest result to the SCUBA counts is obtained for $T_{\mathrm{dust}}=30$ K and $\beta=1.5$. These spectral parameters are consistent with those found from the observations both in the local \citep{dunne} and in the high redshift universe \citep{chapman}, so we will use them hereafter.

The right panel of Figure~\ref{fig:scubatest} shows the differential source counts predicted by the model. As in the plot of integral counts, we show in different colours the contributions of the two active phases of star formation, presented in Figure~\ref{fig:MODEL_phases}. From these plots, it is evident that the short burst phase (blue) dominates the number counts, in particular at the highest fluxes, since it involves a comparable amount of stellar mass over a much shorter characteristic time. The rapid decrease of the differential counts curve in the range $[10-40]$ mJy reflects the rapid change in the merging rate occuring at $z\ga0.7$ and possibly has the same nature as the rapid cosmological evolution of AGN.

\begin{figure*}
\centering
  \includegraphics[width=8.2cm]{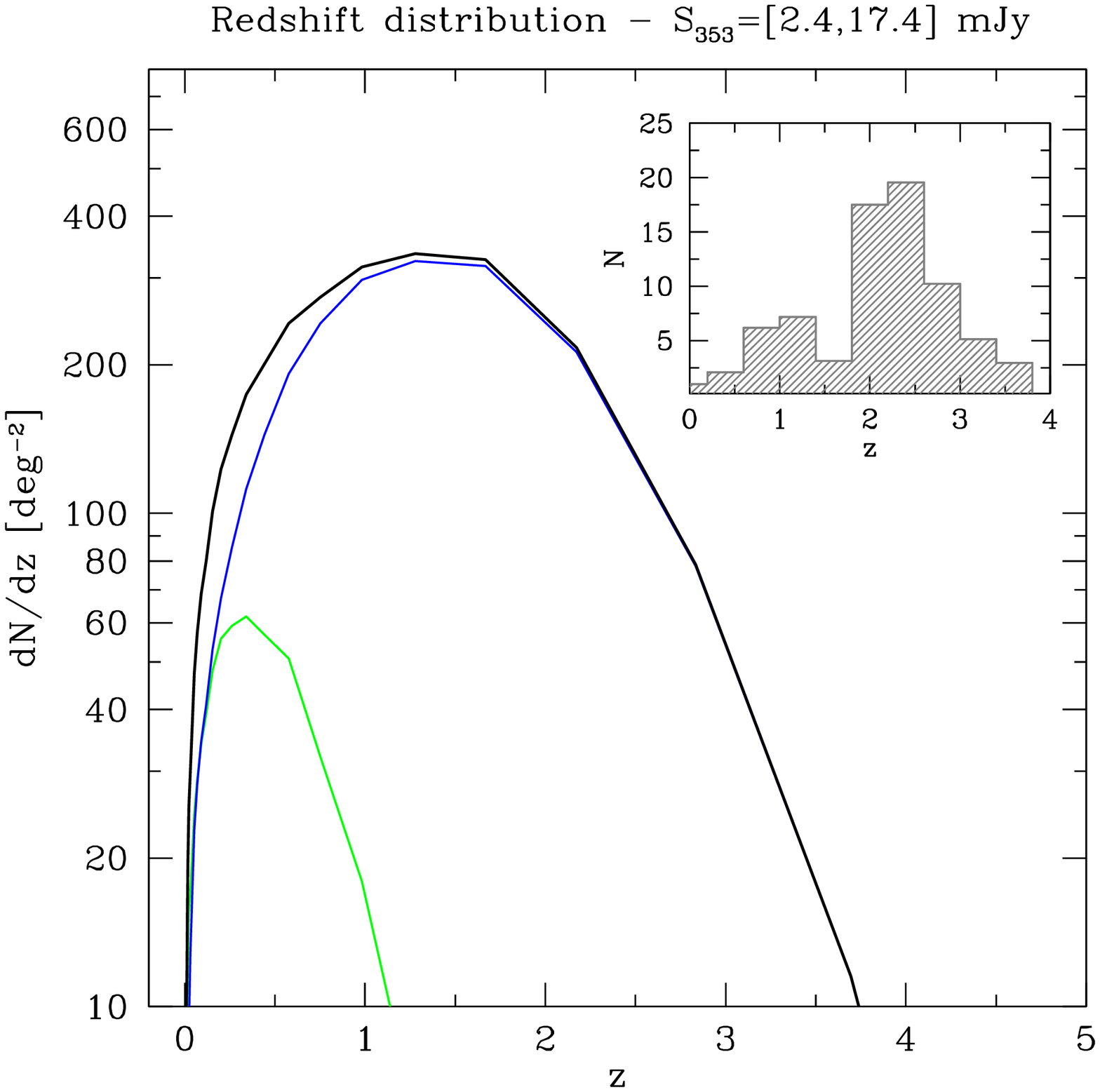}
  \includegraphics[width=8.2cm]{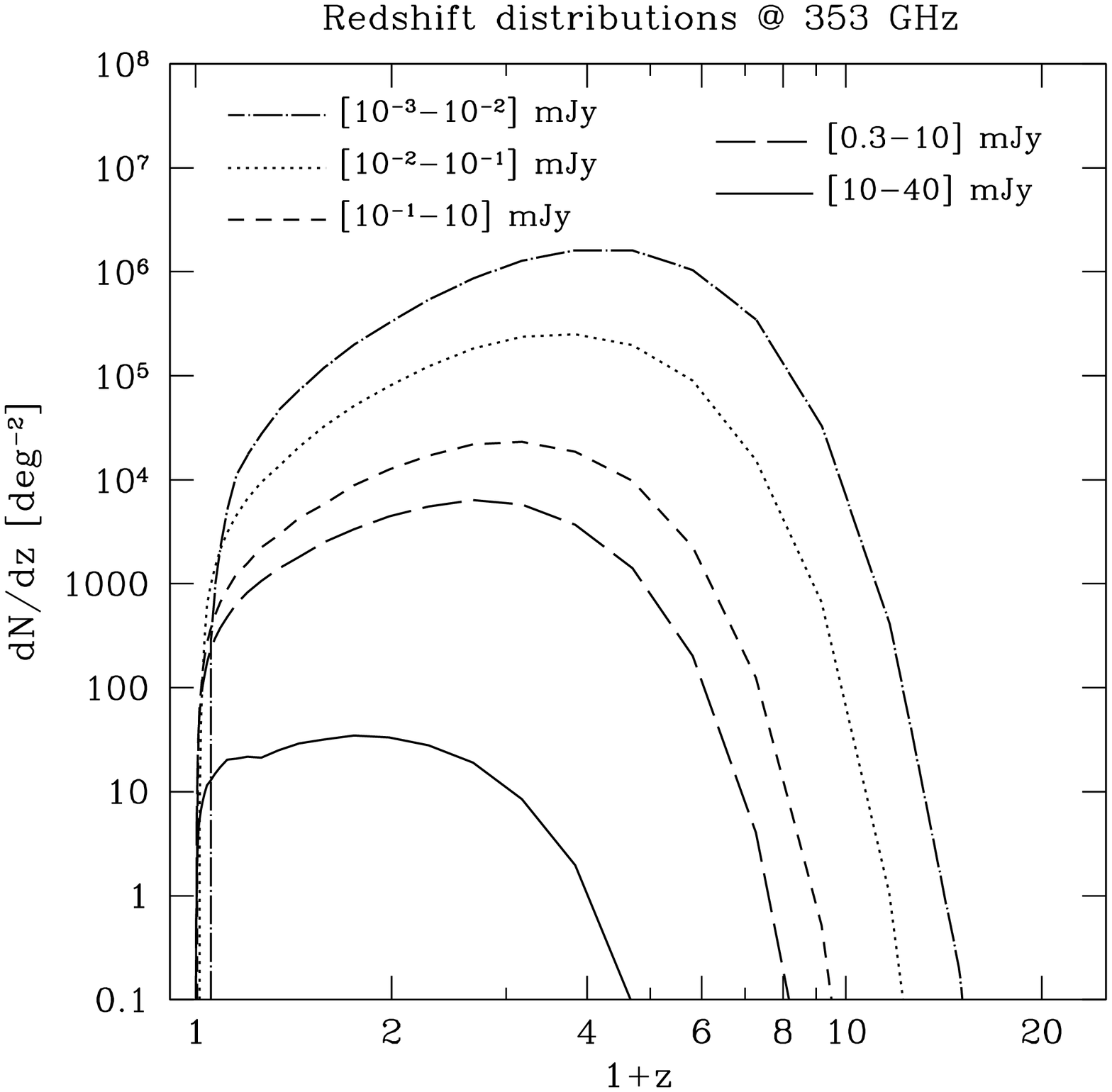}
  \caption{\emph{Left}: The redshift distribution of the sources at 353~GHz according to our model (solid line) for the same flux range of \citet{chapman} and compared with their sample in the insert. Green and blue curves represent, as in Figure~\ref{fig:scubatest}, the contribution of the two star formation phases: also in this case the short phase is dominating and can be detected at much higher redshift. \emph{Right}: the redshift distribution of the sources at the same frequency for different flux ranges: the $[0.3-10]$ mJy interval (long-dashed line) could be relevant for SPT observations, the sources in the range $[10^{-2}-10^{-1}]$ mJy (dotted line) contributes $30\%$ of the sub-millimeter background intensity at 353~GHz.}
  \label{fig:zdistr}
\end{figure*}

As an additional test of these fits, we plot in Figure~\ref{fig:zdistr} the distribution of the sources with redshift. In the left panel we compare with the observations of \citet{chapman}, in the same flux range $[2.4,17.4]$ mJy at 353~GHz. The two distributions are significantly different, in particular our model predicts a peak around $z\sim1.5$ while optical identifications of SCUBA sources give a peak at $z\sim2.5$. This is an intrinsic limitation of the merging models, which predicts the formation of the most massive (and thus brightest) objects at redshifts lower than the redshift range of SCUBA sources. The short burst phase (blue line in the plot) plays the most important role, since it allows the detection of bright sources at much greater distances. The right panel shows several redshift distributions at the same frequency, for different flux ranges.

Note, however, that the source number counts predicted by our model are slightly biased downwards when compared with SCUBA low flux data inferred from lensing
measurements. We have found that, for the small sky area surveyed by SCUBA and
the typical angular correlation scales for sub-millimeter sources, the source counts should show a typical dispersion much greater than the Poissonian estimate. In particular, we have found that typical fluctuations of the order of $20-25\%$ of the {\it real} underlying number source count should be given when trying to estimate this quantity in areas comparable to that covered by SCUBA.

\begin{figure}
  \resizebox{\hsize}{!}{\includegraphics{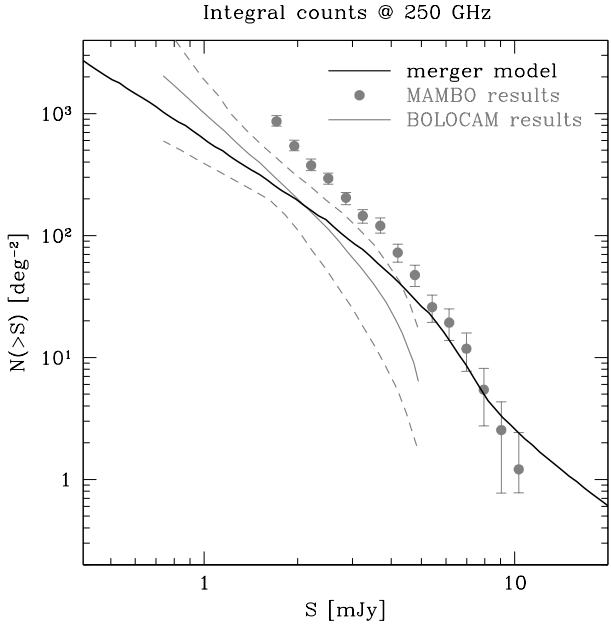}}
  \caption{The integral counts predicted by the merging model at 250~GHz compared with the results of MAMBO \citep{bertoldi} and BOLOCAM \citep{laurent} at the same frequency. The points for BOLOCAM, originally presented at 1.1 mm (273~GHz), have been rescaled to 250~GHz.}
  \label{fig:mambolocam}
\end{figure}

Observations and source counts in the submillimeter band have been also obtained by the MAMBO \citep{bertoldi} and BOLOCAM \citep{laurent} teams at 1.2~mm (250~GHz) and 1.1~mm (273~GHz), respectively. We compared their data with our predictions in Figure~\ref{fig:mambolocam}: we rescaled the BOLOCAM points to 250~GHz using the greybody spectrum assumed in this paper. The total sky area covered by these two instruments is comparable to that of SCUBA. The data of these three experiments are all consistent with each other.

Most of the present and future experiments have a limited angular resolution (around 1 arcmin), which will not permit the resolution of two merging galaxies. The characteristic physical separation of two objects at their first passage is of the order of 100 kpc (corresponding to 12 arcsec at $z\sim2$) and of 10 kpc during the final coalescence (1.2 arcsec). But the impressive sensitivity and angular resolution of ALMA, which is expected to reach a value lower than 1 arcsec in the frequency bands discussed here \citep{alma}, should enable both the identification of merging galaxies as double or asymmetric sources and the distinction between the two phases of star formation activity. Furthermore, ALMA will be able to demonstrate the difference between the typical merging starburst activity and the strong nuclear brightening due to the presence of a hidden AGN.

\subsection{Intensity of the cosmic infrared background and comparison with COBE/FIRAS results}
Another test that can be performed to calibrate this model is to compute the mean brightness of the cosmic infrared background (CIB) due to the point sources and to compare it with the observations at the same frequency. \citet{fixsen98}, on the basis of the observations with COBE/FIRAS, \\
presented the following fit for the intensity of the far-infrared background in the spectral range [150-2400]~GHz
\begin{equation}\label{cibfit}
I_{\nu}^{\mathrm{CIB}}=(1.3\pm0.4)\times10^{-5}(\nu/\nu_0)^{0.64\pm0.12}B_{\nu}(18.5\pm1.2\mbox{ K}),
\end{equation}
where $B_{\nu}$ is the Planck function and $\nu_0$ corresponds to $100\,\mu$m.\\
In our model we estimate the contribution of merging galaxies to the infrared background integrating the source counts over the flux
\begin{equation}
I_{\nu}^{\mathrm{CIB}}=\int S_{\nu}\frac{dN}{dS_{\nu}}dS_{\nu}.
\end{equation}
We predict a large number of sources even at very small fluxes ($S_{353}\sim 10^{-5}$ mJy) and the slope of the differential counts in the range $[10^{-3}-10^{-5}]$ mJy is so steep that these sources do not contribute much to the infrared background intensity. The bulk of it comes from the sources with flux in the range $[10^{-2}-10]$ mJy. However this is strongly related to the assumed value of the escape fraction of photons which strongly diminishes the contribution of less massive objects.

\subsubsection{Escape fraction}\label{subsubsec:fesc}
There is a natural process that decreases the contribution of star-forming galaxies to the intensity of the sub-millimeter background, namely the escape fraction of ionizing photons. Dust inside the haloes absorbes UV and optical photons and then re-emits them in the infrared band. However, assuming that all the ionizing radiation produced inside the galaxies is absorbed by the medium (i.e. by dust), so that the escape fraction is close to zero, is not completely correct, expecially for less massive objects, for which the escape fraction could be greater than zero.\\
We model the escape fraction as a function of the halo mass
\begin{equation}\label{fesc}
f_{\mathrm{esc}}(M)=0.8\,e^{-M/M_{\mathrm{s}}},
\end{equation}
where $M_{\mathrm{s}}$ is the scaling mass, allowing an escape fraction of up to $80\%$ for small-mass objects. To satisfy the limits imposed by the intensity of the infrared background we require a value $M_{\mathrm{s}}=10^{10}\,h^{-1}M_{\odot}$.

\begin{figure}
  \resizebox{\hsize}{!}{\includegraphics{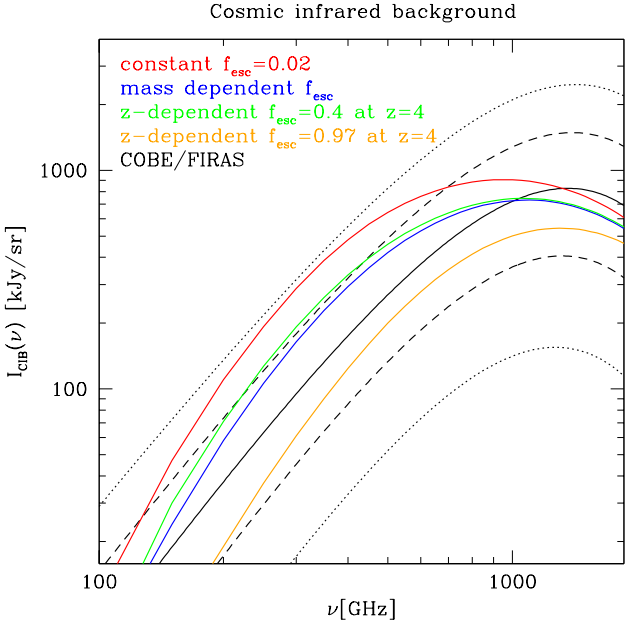}}
  \caption{The spectrum of the cosmic infrared background predicted by our model with a mass-dependent escape fraction (blue), Equation~(\ref{fesc}), compared with the fit to the COBE/FIRAS observations (black) by \citet{fixsen98}, Equation~(\ref{cibfit}): dashed and dotted lines are the $\pm1\sigma$ and $\pm2\sigma$ uncertainty levels, respectively. Red, green and orange lines corresponds to three alternative parametrizations for the escape fraction, see Section~\ref{subsec:fescdep} for details.}
  \label{fig:firas}
\end{figure}

\begin{figure}
  \resizebox{\hsize}{!}{\includegraphics{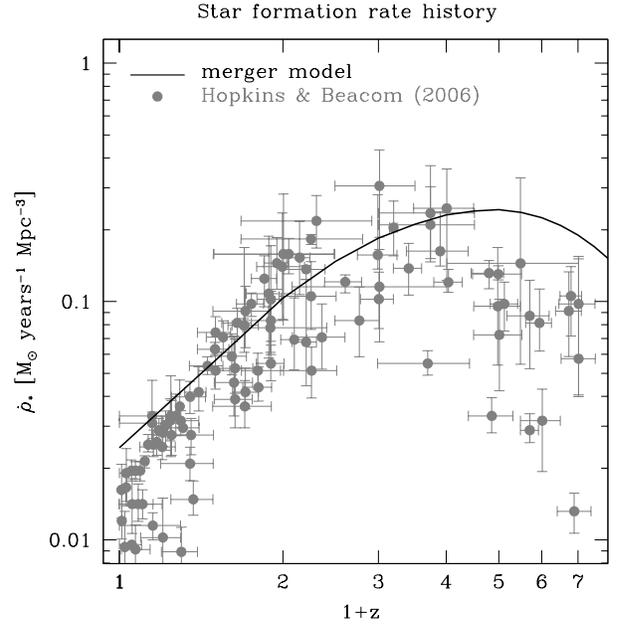}}
  \caption{The evolution of the star formation rate according to the merger model (solid line), compared with the compilation of \citet{hopkins06} of numerous observations at different wavelengths (dots).}
  \label{fig:sfr}
\end{figure}

The idea of a mass-dependent escape fraction was used in the works of \citet{escfrac1} and \citet{escfrac2}, which dealt with the reionization of the universe. Here we are more concerned with the presence of dust which absorbs ultraviolet photons and then influences the calibration of the Kennicutt relation, which assumes that almost all photons produced by young stars are absorbed by dust and re-emitted in the infrared band. In this sense, we need to compare the infrared and ultraviolet luminosities of observed galaxies to derive an estimate of the escape fraction in these objects.\\
For example, very recent observations by \citet{buat} using SPITZER and GALEX allow us to estimate an escape fraction of the order of $1-2\%$ for bright galaxies in the local universe. A very tight correlation between the $L_{\mathrm{TIR}}/L_{\mathrm{FUV}}$ ratio\footnote{TIR indicates the $[1-1000]$~$\mu$m band} and metallicity has been found by \citet{cortese} in a sample of nearby normal star-forming galaxies: the infrared to ultraviolet ratio increases with metallicity, thus leading to a higher escape fraction for low metallicity objects. Earlier observations by \citet{heckman} seem to confirm this trend even for starburst galaxies, with escape fractions of the order of $30-50\%$ for very metal poor objects.
The metallicity is also known to correlate very well with stellar mass, as shown by \citet{tremonti} and \citet{erb} for the local and the high-redshift universe, respectively. In the merger model presented here stellar mass inside every halo is directly related to the mass of the halo itself. Then it is natural to expect a higher metallicity in more massive systems and therefore also a lower escape fraction. These simple arguments demonstrate that modelling this quantity as a function of mass is, at least qualitatively, well justified.

As mentioned before, the Kennicutt relation implicitly assumes that almost all the ultraviolet photons produced by young stars are absorbed and reradiated by dust in the infrared, i.e. that $f_{\mathrm{esc}}=0$. When considering non-zero values of $f_{\mathrm{esc}}$, we need to introduce a correction in the Kennicutt relation
\begin{equation}\label{kenniccorr}
\dot{M}_{\star}\,[M_{\odot}/\mathrm{yr}]=\frac{1.71\times10^{-10}}{1-f_{\mathrm{esc}}}\,L_{\mathrm{IR}}\,[L_{\odot}].
\end{equation}
In this way, less massive objects are unable to retain all their ionizing radiation, dust absorption will be less efficient and this results in a lower infrared luminosity for a given star formation rate. We will return to this point in Section~\ref{subsec:fescdep}, discussing several alternative parametrizations and their effect on the CMB fluctuations.\\
As shown in Figure~\ref{fig:firas}, with such a correction, our prediction for the intensity of the cosmic infrared background is within the 1$\sigma$ limits of the fit in Equation~(\ref{cibfit}). On the other hand, with an escape fraction independent of mass and redshift and close to the $2\%$ value of \citet{buat}, our model predicts an intensity for the sub-millimeter background which exceeds the COBE/FIRAS value but remains within the $2\sigma$ errorbars.

About $30\%$ of the background intensity at 353~GHz comes from sources in the flux range $S_{353}=[0.01-0.1]$ mJy. The redshift distribution of such sources is displayed by the dotted line in the right panel of Figure~\ref{fig:zdistr} and it is characterized by a broad peak around $z\simeq2.5-3$.

\subsection{The Madau plot for cosmic star formation history}\label{subsec:madau}
The very simple parametrization which we use to model the star formation process within the framework of the halo model can be verified computing the cosmic star formation density as a function of redshift \citep[i.e. the \emph{Madau plot},][]{madau96,madau98,hopkins04,hopkins06} and comparing it with the observations.

We average the star formation rate over all haloes $M_1$ which are merging with other haloes to produce a final mass $M$
\begin{equation}\label{dotrho}
\dot{\rho}_{\star}(z)=\int dM_1 \int dM \frac{dN_{\mathrm{merg}}}{dMdt}\frac{dn}{dM_1}\,M_{\star}(M_1,M_2,z),
\end{equation}
where $\frac{dN_{\mathrm{merg}}}{dMdt}$ is defined in Equation~(\ref{LC}) and $M_{\star}(M_1,M_2,z)$ is the number of stars produced in each merger episode.

The evolution of cosmic star formation rate density is compared with recent observations, obtained at different wavelengths, in Figure~\ref{fig:sfr}. It is clear that our model slightly overestimates the cosmic star formation rate, especially at higher redshift ($z\ga5$), where our points are outside the observational errorbars. This could be due again to the limitations of the merging model which, though very accurate at the lower redshifts, fails to reproduce the merger trees at high redshift.\\

\noindent Using a compromise model that is consistent with the measurements of the cosmic star formation rate that does not exceed the SCUBA number counts and that is close to the 1$\sigma$ limit of the sub-millimeter background provided by COBE/FIRAS, we can make predictions for the fluctuations in the CMB connected with the foreground sub-millimeter emission from star forming galaxies.

\begin{figure}
\centering
   \resizebox{\hsize}{!}{\includegraphics{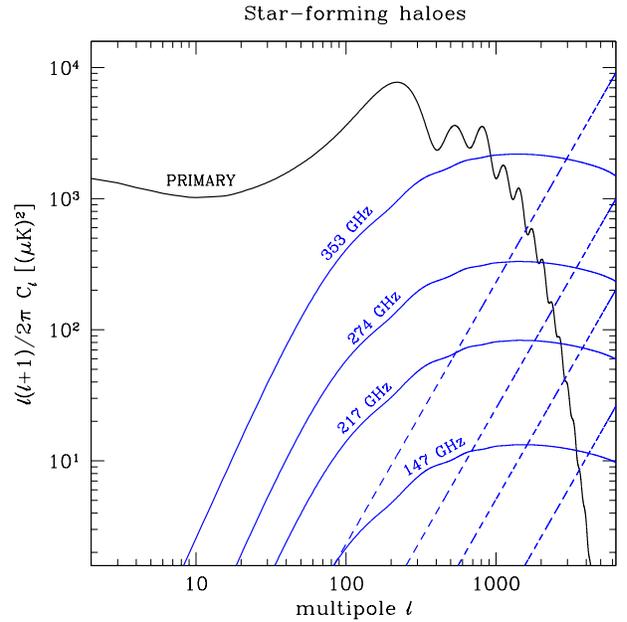}}
  \caption{The Poisson (blue dashed line) and correlation (blue solid line) term of dust emission inside star-forming haloes, compared with the primary signal (black), at the four frequencies of the future CMB experiments (PLANCK, SPT and ACT): 353, 274, 217 and 147~GHz. Sources brighter than 100 mJy at 353~GHz have been removed in the computation of Poisson term.}
  \label{fig:clhalo}
\end{figure}

\begin{figure}
  \resizebox{\hsize}{!}{\includegraphics{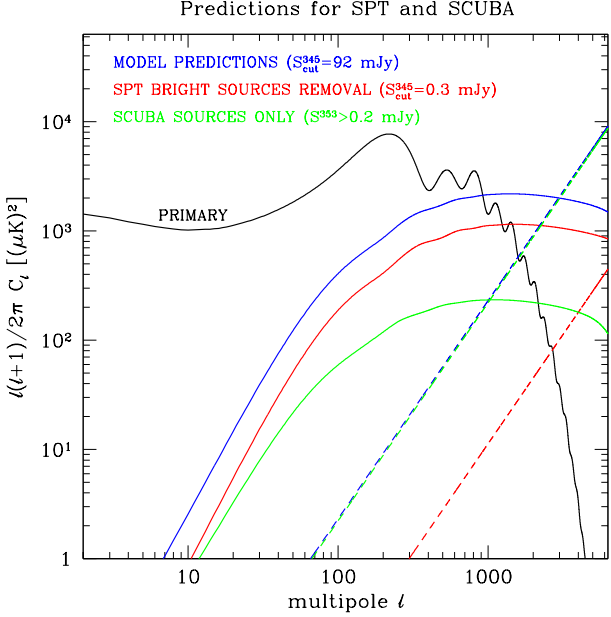}}
  \caption{The angular power spectrum of star-forming haloes at 345~GHz: the red line is the predicted level of the fluctuations after the removal of the sources brighter than $\sim0.3$ mJy \citep[corresponding to SPT sensitivity,][]{spt}: compared with the model (blue) the Poisson term is reduced by more than one order of magnitude and the correlation term, which is much less dependent on bright sources, dominates up to higher multipoles. The green line is the predicted power spectrum considering only the sources detectable by SCUBA (with lensing technique) at 353~GHz (i.e. with flux greater than 0.2 mJy): the contribution of such sources to the correlation term is practically insignificant, while they make all the Poisson power.}
  \label{fig:clhaloSPT}
\end{figure}

\section{Results}\label{sec:result}
In this section we present the angular power spectrum resulting from the distribution of dusty sources in the universe. As shown in Appendix~\ref{app:powerspec}, such a power spectrum is the sum of two terms, the Poisson $C_l^{\mathrm{P}}$ and the correlation $C_l^{\mathrm{C}}$ 
\begin{equation}
\langle\Delta I_{\nu}(\vec{n}_1)\Delta I_{\nu}(\vec{n}_2 )\rangle=\sum_l\frac{2l+1}{4\pi}\left(C_l^{\mathrm{P}}+C_l^{\mathrm{C}}\right)P_l(\vec{n}_1 \cdot \vec{n}_2),
\end{equation}
where $l$ is the multipole index and $P_l$ is the Legendre polynomial of order $l$.\\
The Poisson contribution to the angular power spectrum can be obtained from the source number counts \citep[e.g.][]{scottwhite}
\begin{equation}\label{poisson}
C_l^{\mathrm{P}}(\nu)=\int_0^{S_{\mathrm{cut}}}S_{\nu}^2\frac{dN}{dS_{\nu}}dS_{\nu}
\end{equation}
where all the sources brighter of $S_{\mathrm{cut}}$ are supposed to be removed. We present the results for $S_{\mathrm{cut}}=100$ mJy at 353~GHz, unless otherwise specified.

The power spectra generated according to Equation~(\ref{poisson}) and by (\ref{correl}) for the correlation term are shown in Figure~\ref{fig:clhalo} at the four different frequencies of the future CMB experiments. The correlation term peaks in a broad range $l=[200-3000]$. We notice that the Poisson signal is smaller than the correlation one for $l\la3000$ and becomes dominant only at the very large multipoles. Such multipoles, corresponding to angular scales of the order of a arcminute, are within the resolution limits of ACT and SPT. This plot shows that even at 147~GHz the fluctuations due to the dust emission in merging galaxies are one of the most important detectable extragalactic foregrounds. At higher frequencies it becomes dominant and at 353~GHz it should exceed the contribution from higher peaks (beginning from the fourth). As is well known, at 217~GHz the signal of the SZ clusters is negligibly small and this makes it much easier to observe the contribution from star-forming haloes.

\begin{figure}
  \resizebox{\hsize}{!}{\includegraphics{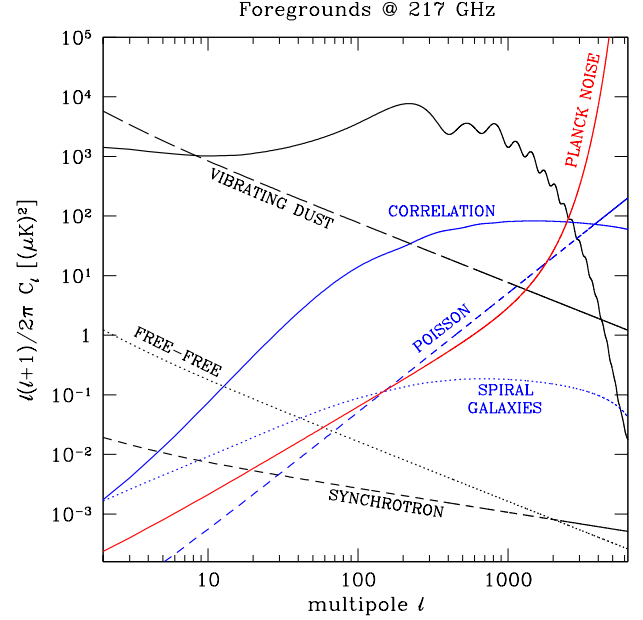}}
  \caption{The effect of foregrounds according to the model of \citet{tegmark}, at 217~GHz: the long-dashed line refers to vibrating dust, free-free emission is plotted with the dotted-line and synchrotron radiation with the short-dashed line. The uncertainty on the noise level at 217~GHz for PLANCK is the red solid line, for a sky coverage $f_{\mathrm{sky}}=85\%$. Solid and dashed blue curves represent our results for star-forming haloes, the dotted curve is the correlation term from spiral galaxies (see Section~\ref{subsec:spir}).}
  \label{fig:clforegr}
\end{figure}

Since Poisson power is generated mainly by very bright sources ($S\ga1$ mJy), future experiments should be able to detect such sources in the maps and eventually to remove them. In this way the correlation term can be detected also at very small scales, since it is slightly affected by rare bright sources and obtains the bulk of its power from the more abundant and less bright sources. We estimate this in Figure~\ref{fig:clhaloSPT} at the 345~GHz SPT channel. This instrument is expected to have a sensitivity of about 0.3 mJy, and this should allow one to excise the brightest sources and hence increase the weight of the correlation term versus the Poissonian one. In the same plot we also compare the predictions of our model with the result that would be obtained if only the sources detectable by SCUBA were considered. The Poisson term would be at the same level, since it is generated mainly by the bright objects. The correlation would be strongly reduced, since SCUBA is not able to detect sources with flux lower than 0.2 mJy, missing the abundant dim sources which contribute significantly to the correlation power.

\begin{figure*}
\centering
  \includegraphics[width=8.2cm]{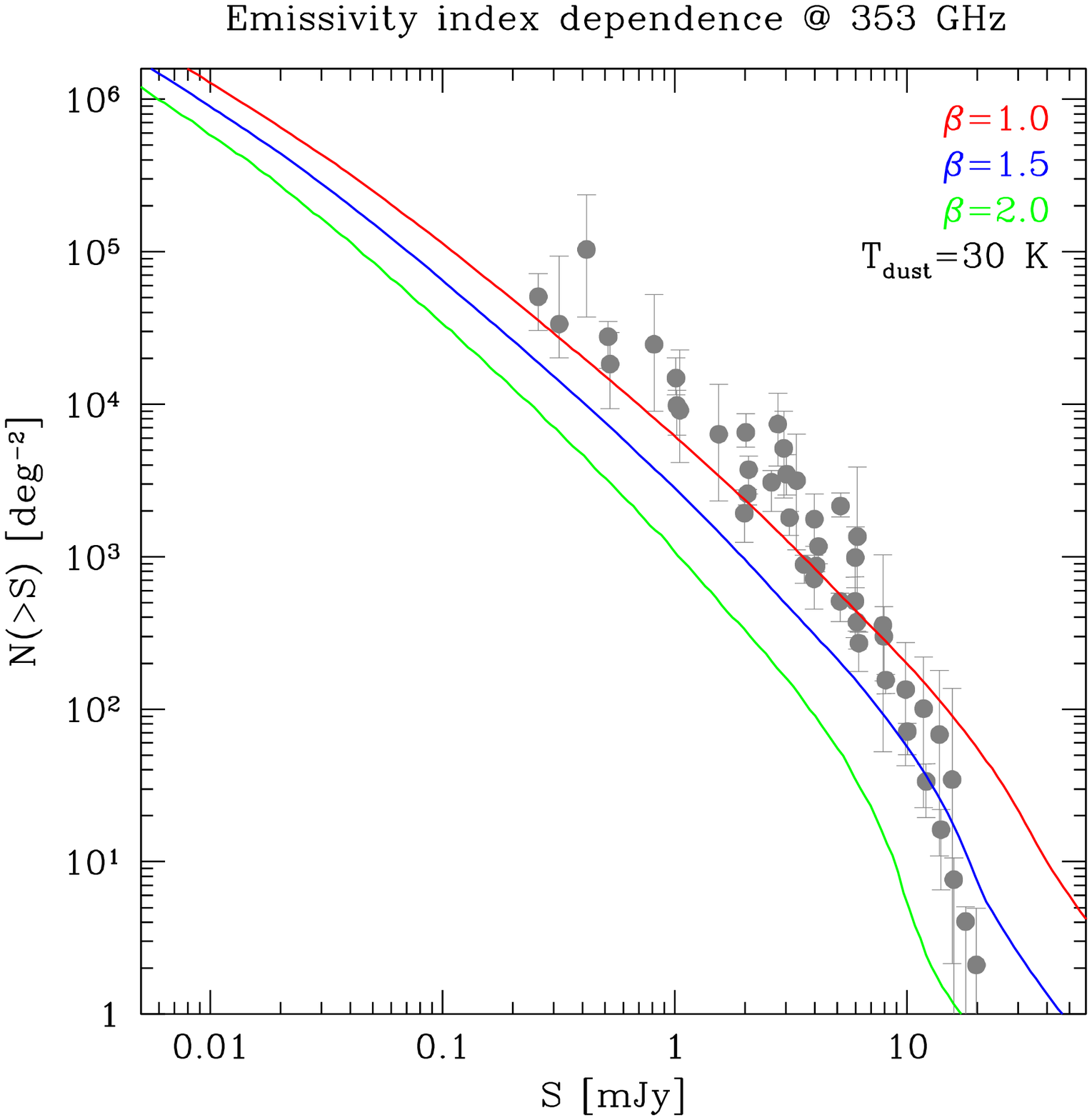}
  \includegraphics[width=8.2cm]{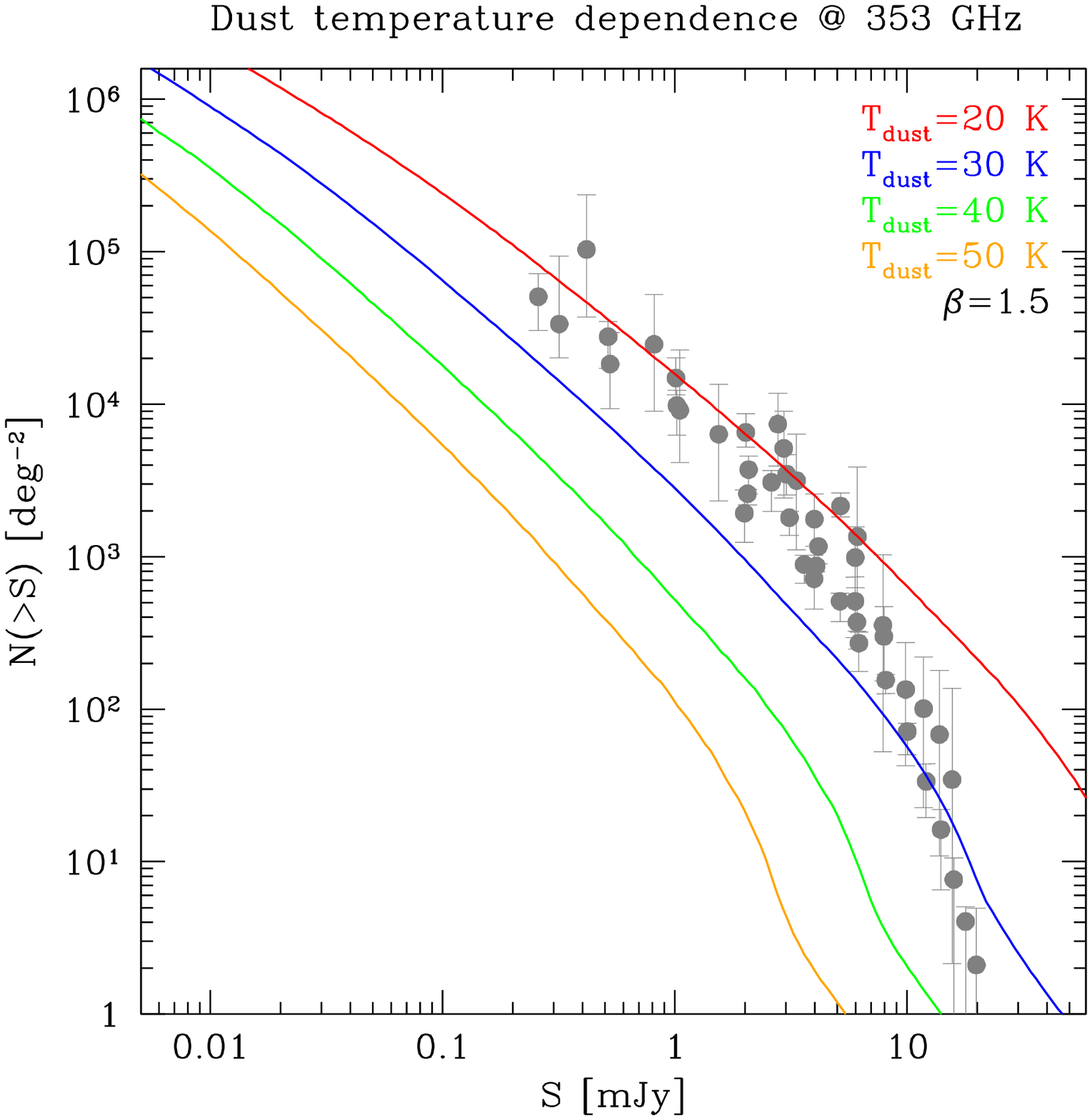}
  \caption{Sensitivity of the results for the integral source counts at 353~GHz to the choice of the spectral parameters: emissivity index (\emph{left}, for fixed value of the dust temperature, 30 K) and dust temperature (\emph{right}, for fixed value of the emissivity index, 1.5). Grey dots are the SCUBA observations, as in Figure~\ref{fig:scubatest}. The curves in blue represent the choice which we use for all the results presented in the paper. The plausible range of dust temperatures derived from the observations \citep{dunne,chapman} is 30-50 K.}
  \label{fig:countsparam}
\end{figure*}

The analysis of the differential contribution to the signal from sources at different redshift reveals that the bulk of the correlation is generated by star forming haloes in the redshift range $z\simeq2-4$, with a slight dependence on the multipole $l$.

Finally, following \citet{ik04}, we estimate the amount of intergalactic dust expelled from galaxies by strong winds and AGN. This dust is heated by the ultraviolet background and by the CMB. As shown by \citet{basu}, its effect on the CMB angular power spectrum will be $\delta C_l \simeq -2\tau_{\mathrm{dust}}(\nu) C_l$ (high $l$), where $C_l$ is the primary signal and $\tau_{\mathrm{dust}}$ is the absorption optical depth. This turns out to be three orders of magnitude below the signature of the dusty merging haloes discussed in this paper.

\subsection{Contribution from normal spiral galaxies}\label{subsec:spir}
Another contribution comes from the infrared emission of normal spiral galaxies, like the Milky Way. We decribe the distribution of such objects in space with the Press-Schechter mass function, weighted with the halo occupation number which gives the number of (blue) galaxies in a halo of given mass. \citet{spnum} propose the following form for such a quantity
\begin{equation}
N_{\mathrm{gal}}(M)=\left(\frac{M}{M_{\mathrm{gal}}}\right)^{\alpha}+0.5 \exp\left\{-4\left[\log_{10}(M/10^{11.75})\right]^2\right\},
\end{equation}
where $M$ is the mass of the halo and with $M_{\mathrm{gal}}=7\times10^{13}\,h^{-1}M_{\odot}$ and $\alpha=0.9$. This fit is calibrated on the results of semi-analytical models and does not take into account the dependence on redshift.\\
The correlation term resulting from this simple estimate is shown in Figure~\ref{fig:clforegr} (dotted curve) and turns out to be more than two orders of magnitude smaller than the signal due to the star-forming haloes.

\begin{figure*}
\centering
  \includegraphics[width=8.2cm]{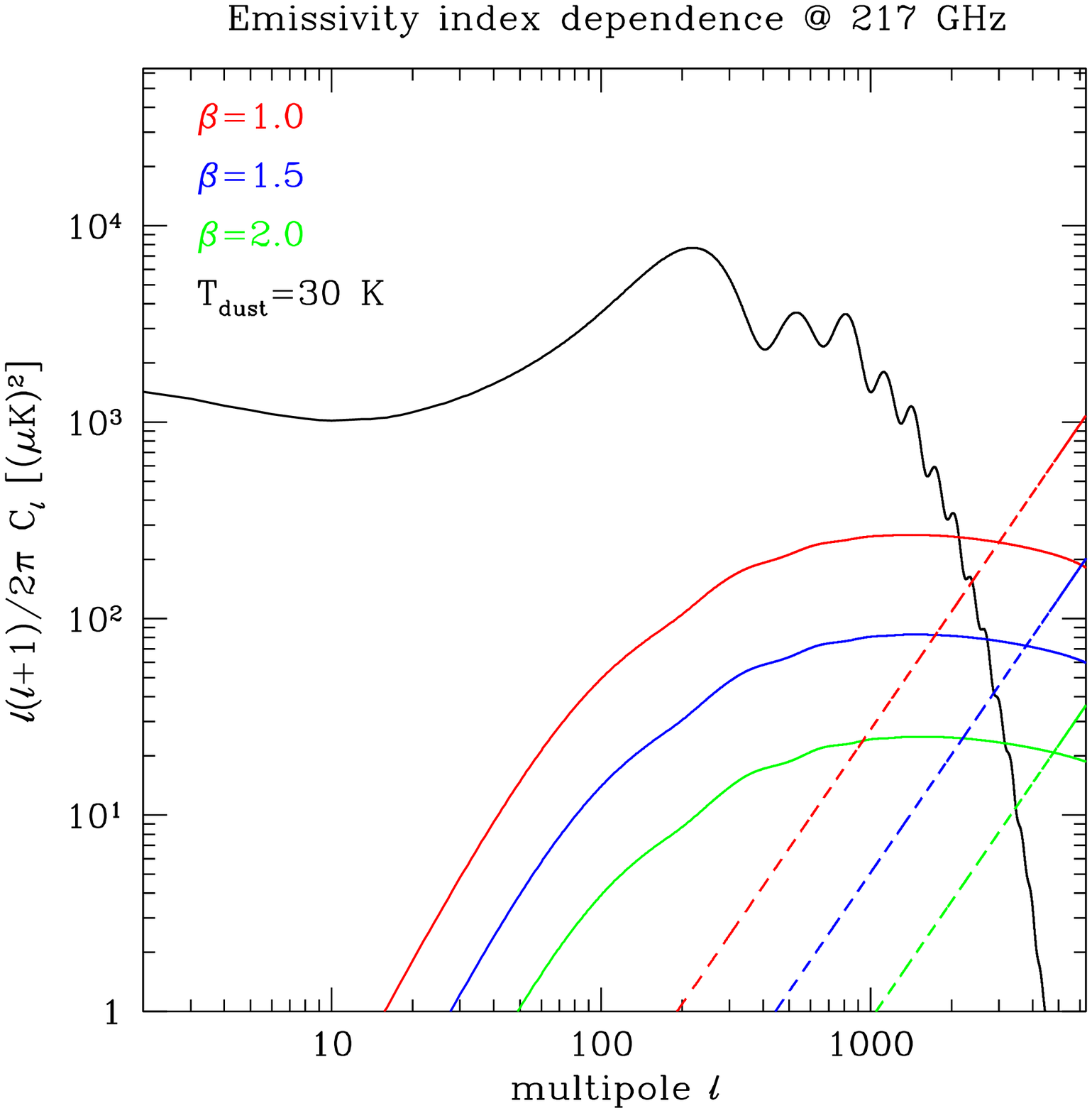}
  \includegraphics[width=8.2cm]{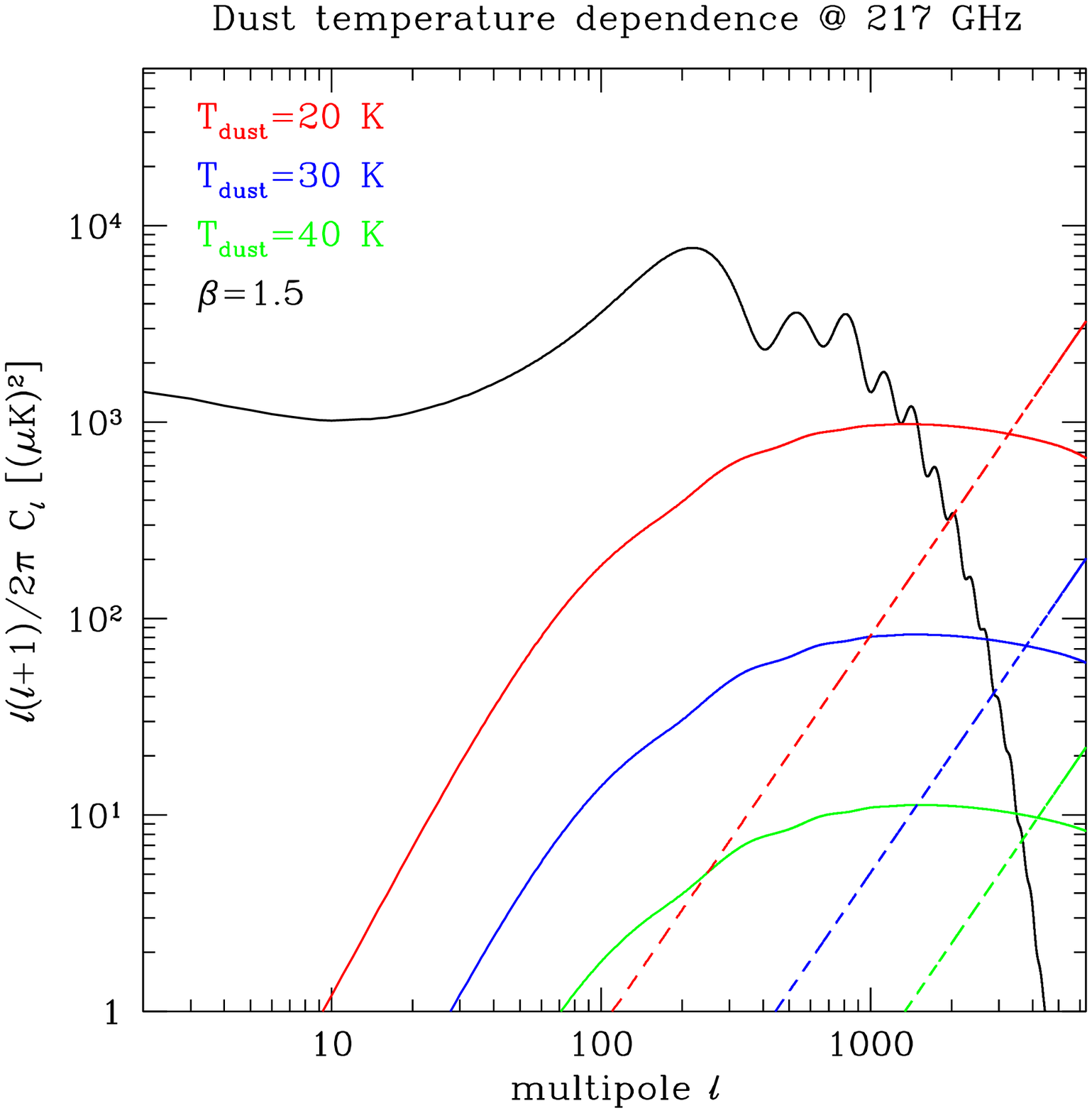}
  \caption{Sensitivity of the Poisson and correlation term at 217~GHz to the choice of the free parameters for the dust spectrum: emissivity index $\beta$ (\emph{left}, for fixed value of the dust temperature, 30 K) and dust temperature $T_{\mathrm{dust}}$ (\emph{right}, for fixed value of the emissivity index, 1.5). The curves in blue represent the choice that we use for all the results presented in the paper.} 
  \label{fig:clparam}
\end{figure*}

\subsection{Comparison with galactic foregrounds sources}
In Figure~\ref{fig:clforegr} we present the well-known model of \citet{tegmark} for the main galactic foregrounds at 217~GHz (since at this frequency the contribution from the thermal SZ is zero): the most intensive signal comes from the dust emission in our galaxy (long-dashed line), but it decreases rapidly at larger $l$, where our effect becomes dominant. The contribution from free-free emission (dotted line) and synchrotron radiation (short-dashed line) have similar dependence on $l$, but with much less power and should not affect the detection of the correlation term of dust in star-forming regions.\\
We also include the expected uncertainty of the noise level for the PLANCK HFI detectors. At 217~GHz the expected resolution is 5 arcmin, with a sensitivity $\Delta T/T=4.8\times10^{-6}$. ACT and SPT telescopes are expected to give even better values of both resolution and sensitivity and hence a much lower level of noise. This should allow the detection of the dusty sources in the power spectrum.

\subsection{Comparison with previous works}
An estimate of the expected angular power spectrum of the cosmic infrared background was given by \citet{haikno}. The model that they were using is different from the one presented here: it assumes a Press-Schechter distributed population of sources with a mass and redshift dependent dust content and temperature related to the stellar field. Our model instead considers the distribution of \emph{merging} objects, derived from the \citet{lc93} approach, and thus directly probes the star forming haloes. Given these different approaches, we compare our result at 353~GHz with their ``hot dust'' model and find a rather good agreement: our correlation term at $l=1000$ is a factor $\sim1.4$ (in $\Delta T$) lower than their estimate.

\citet{song} performed detailed computations for the FIR background, modelling the dust emission from star-forming galaxies within the framework of the halo model, but not considering merging. They presented the angular power spectra for the 545~GHz channel of PLANCK assuming a sky coverage corresponding to the cleanest 10\% and including the unsubtracted Galactic dust. Their results are essentially in agreement with \citet{haikno}.

More recently, \citet{gonznuevo} presented their predictions for the angular power spectrum of extragalactic point sources, based on the theoretical model of \citet{granato}. Comparing with their result at 353~GHz we find that at $l=1000$ their correlation and Poisson terms are higher by a factor $\sim2.8$ and $\sim1.6$, respectively, than our predictions.

\citet{toffolatti} presented an estimate of the power spectrum of ACBAR sources at 150~GHz. Our estimate for the Poisson term at the same frequency (and for the same flux limit $S_{\mathrm{cut}}^{150}=24$ mJy) is roughly a factor 1.6 lower; we have a similar deficit when comparing with SCUBA observations.

From these comparisons, it is clear that the contribution to the background fluctuations from the sub-millimeter sources in our model is lower than predicted by other authors, but in general their results are confirmed.

\begin{figure}
  \resizebox{\hsize}{!}{\includegraphics{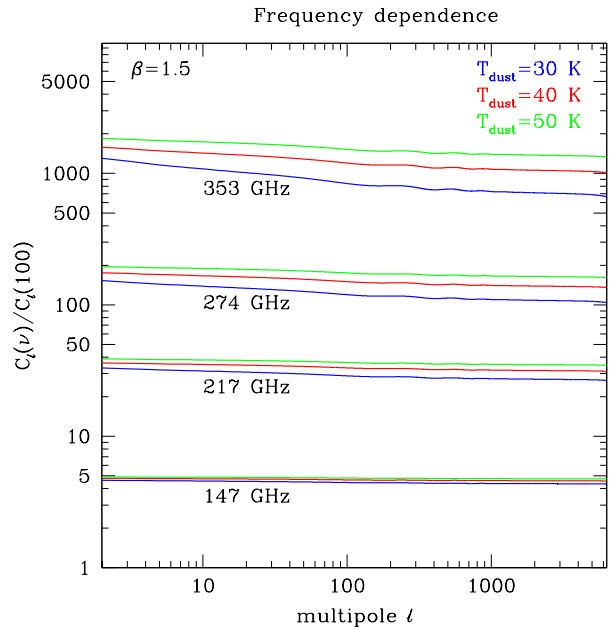}}
  \caption{The ratio of the correlation signal at different frequencies to the signal at 100~GHz, for different values of the dust temperature.}
  \label{fig:clnu}
\end{figure}

\section{Sensitivity to the parameters: can we learn something about the physical nature of the sources?}\label{sec:sens}
The model presented in this paper is sensitive to several parameters. As we pointed out in Section~\ref{sec:test}, it may be feasible to calibrate these parameters with the existing observations.
The main purpose of this work is to make predictions for the level of foreground of dusty sources in the universe and to show that this signal will be important for upcoming experiments.\\
However, this component should open an additional way to study the population of the most distant star-forming galaxies, which are already enriched by dust. The amplitude of the signal could provide a very useful tool to understand the physical nature of the objects. For example, the amount of dust could be derived if estimates are obtained about the actual value of the escape fraction, since these two quantities are related. Multifrequency observations will permit one to separate the contribution of distant dusty haloes from other background/foreground sources due to their characteristic spectrum, and could lead to a more precise estimate of the spectral properties of dust emission, in particular the temperature. The extremely high resolution of ALMA will make it possible even to individually detect very distant objects and to provide hints about their dust emission, in particular if information about their redshift is also retrieved (by observing, for instance, the emission in the fine-structure lines of the most abundant ions which were detected by COBE/FIRAS in the sub-millimeter spectrum of our galaxy and which were discussed in \citet{basu}). Measuring the spectrum of the additional fluctuations in the CMB at different multipoles may also set constraints on the star formation scenario and confirm that the star formation at high  redshift begins in strongly overdense regions which will eventually evolve into clusters and superclusters of galaxies at $z\sim0.5-0.7$ and that the first star-forming haloes very rapidly create a high abundance of metals and dust.\\
In this section we will explore the parameter space and present several results for the angular power spectrum as a function of different parameters.

\begin{figure*}
\centering
  \includegraphics[width=8.2cm]{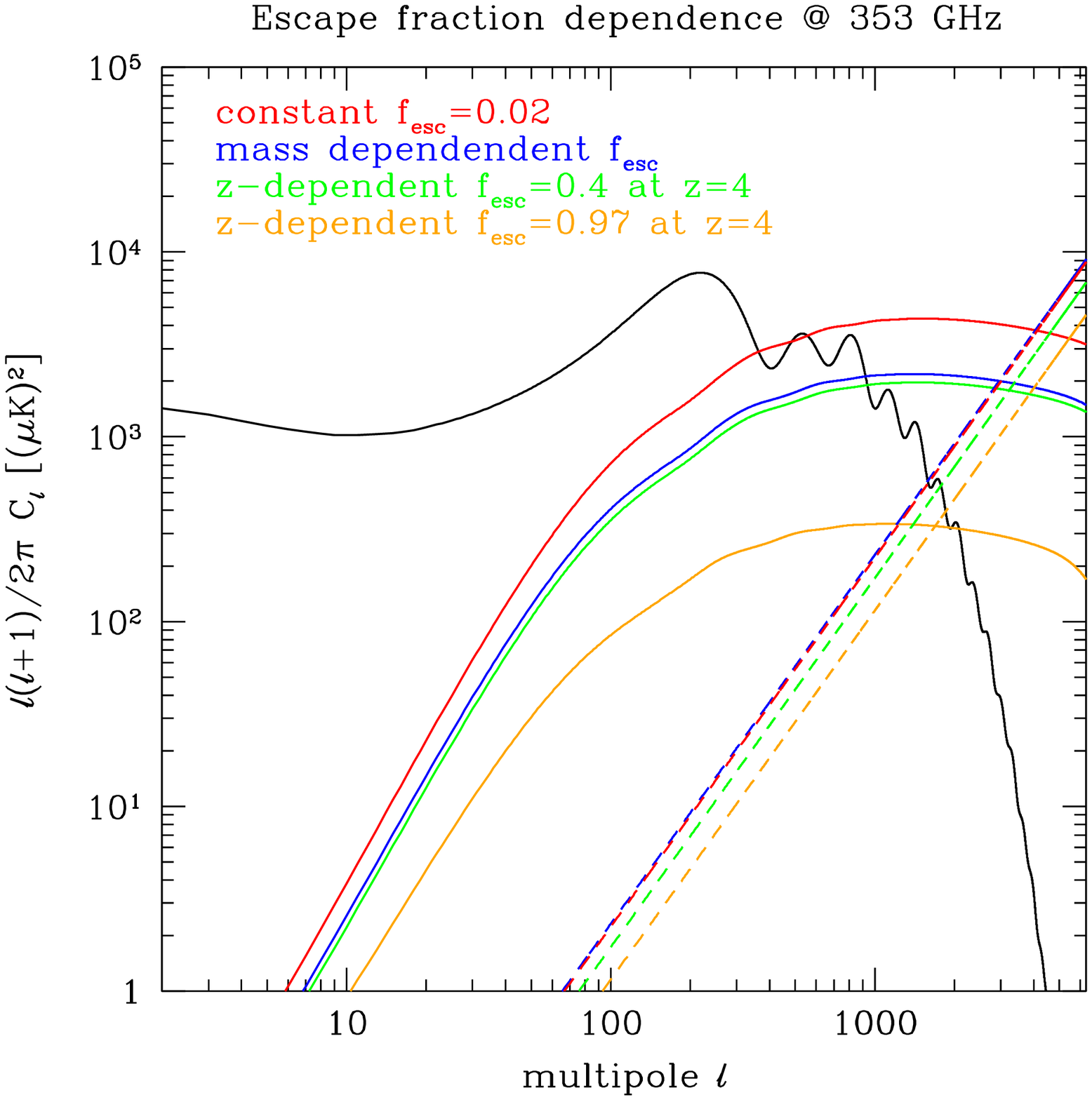}
  \includegraphics[width=8.2cm]{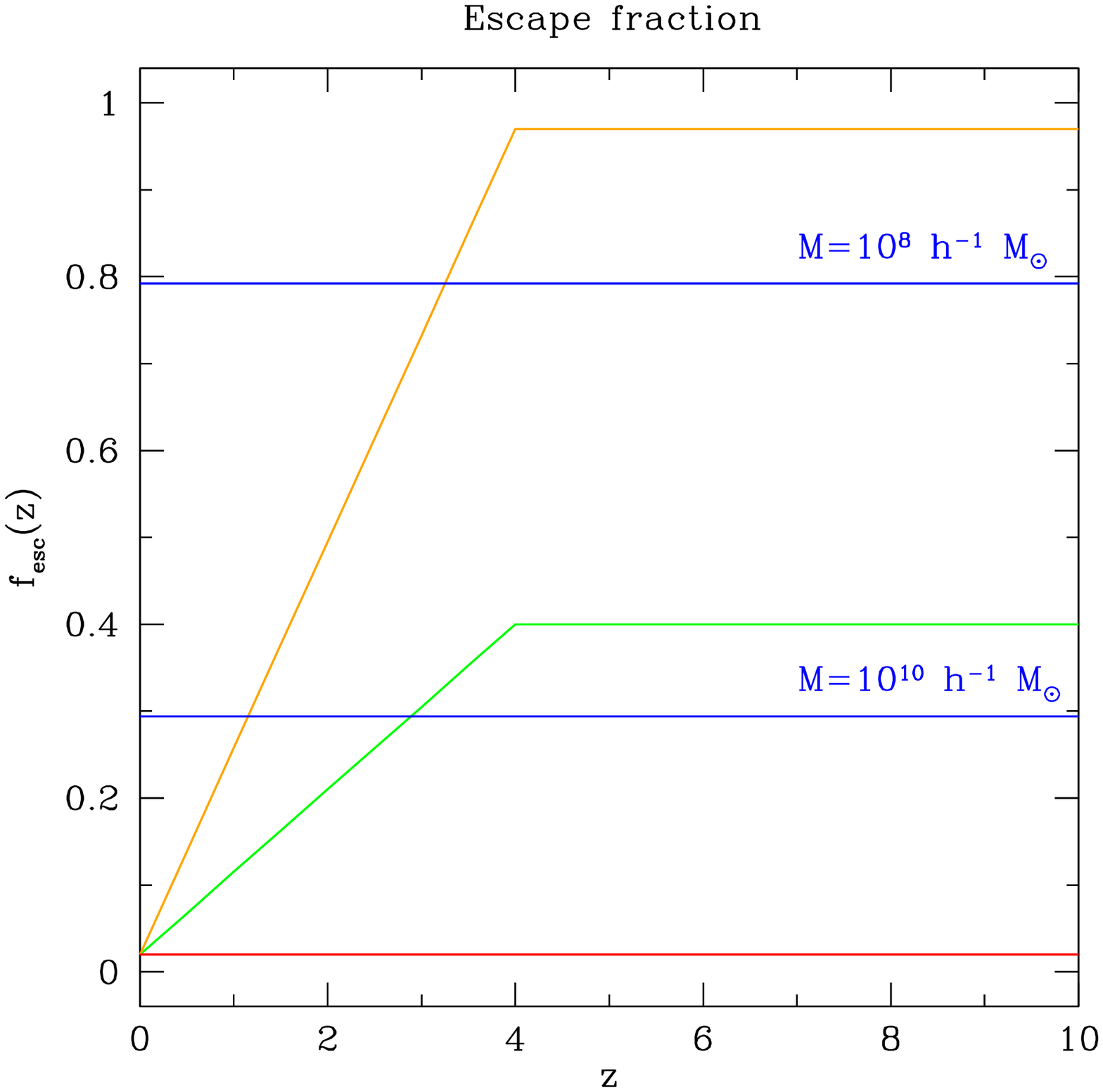}
  \caption{\emph{Left:} the dependence of the correlation (solid) and Poisson (dashed) term on different parametrization for the escape fraction with redshift (\emph{right}) at 353~GHz: low and constant (red), low and redshift dependent (green), high and redshift dependent (orange); blue lines correspond to the mass-dependent escape fraction adopted in the compromise model and show constant redshift evolution for given mass.}
  \label{fig:clfesc}
\end{figure*}

\subsection{Spectral parameters}\label{subsec:sensparam}
The shape of the correlation term is fixed once the merger model is established, since it determines the distribution of star-forming haloes as a function of their luminosity. However, the merger model is connected only to the total infrared luminosity through the Kennicutt relation. To obtain the value of the anisotropy in a specific frequency channel, one has to assume some kind of spectral distribution for the sources. In Section~\ref{sec:dust} we assumed a greybody distribution depending on the emissivity index $\beta$ and on the dust temperature $T_{\mathrm{dust}}$. Based on the comparison with the SCUBA source counts (Figure~\ref{fig:countsparam}), we chose $\beta=1.5$ and $T_{\mathrm{dust}}=30$ K. These values are in agreement with the observational constraints, which give a wide range of values both for the emissivity index ($1.0\le\beta\le2.0$) and the dust temperature ($30\mbox{ K}\le T_{\mathrm{dust}}\le50\mbox{ K}$). As shown in Figure~\ref{fig:clparam}, the amplitudes of the correlation and Poisson term are sensitive to the choice of the spectral parameters and display a variation over one order of magnitude in the range of possible values. It is clear, from these plots, that the two spectral parameters show a degeneracy. Multifrequency observations may allow one to reduce the extent of such degeneracy, as shown in the following.

\subsection{Frequency dependence}
As explained in Section~\ref{sec:dust}, the contribution from high redshift objects in the highest frequency bands of future experiments is expected to be low, due to the shift at wavelengths shorter than the 100~$\mu$m peak in the emission spectrum. On the other hand, moderate frequency channels will be very sensitive even to extremely distant objects, since their emission will occur at wavelengths longer than the peak. This is shown in Figure~\ref{fig:clnu}, where the correlation term, normalized to the signal at 100~GHz, is compared for different frequencies: while this ratio stays almost constant at 147, 217 and 274~GHz, it decreases at 353~GHz, towards the higher multipoles, corresponding to the smallest angular scales and hence to the most distant objects, whose emissivity at these frequencies is far beyond the peak, on the Wien tail of the spectrum.\\
The level of such a decrease at higher frequencies is somewhat temperature dependent. Higher dust temperatures shift the position of the peak in the spectrum towards higher frequencies and shift the position of the sources beyond the peak towards higher redshift. This effect could be useful to measure the dust temperature of extremely distant objects, breaking the degeneracy between $\beta$ and $T_{\mathrm{dust}}$ pointed out above, since the value of $\beta$ has no influence on the position of the peak in the spectrum.

\subsection{Escape fraction}\label{subsec:fescdep}
As we mentioned in Section~\ref{subsubsec:fesc}, we introduce an escape fraction for the ionizing photons in order to satisfy the limits of COBE/FIRAS on the intensity of the infrared background. In the compromise model, which provides all the main results of this work, the escape fraction is modelled as a function of the halo mass (Equation~\ref{fesc}). Here we present some other possibilities and show how these would affect our results on the fluctuations.\\
GALEX and SPITZER recently obtained several estimates for the escape fraction in nearby luminous galaxies \citep[$L_{IR}\ga10^{11}\,L_{\odot}$, see e.g.][]{buat}. The published data do not give precise information about the value of the escape fraction at high redshift ($z\sim3-5$) from where most of the contribution to the correlation term comes. Similarly, up to now there is no clear idea about the dependence of the escape fraction on galaxy mass. We present here three further parametrizations: a constant and low escape fraction of $2\%$, an escape fraction growing with redshift from $2\%$ at $z=0$ to $40\%$ at $z=4$ and another one from $2\%$ to $97\%$. \\
The correlation and Poisson terms at 353~GHz, corresponding to these parametrizations are plotted in the left panel of Figure~\ref{fig:clfesc}: the blue line is the compromise model with the mass dependent escape fraction, the red line is the constant $f_{\mathrm{esc}}=0.02$: the correlation term in this case in much stronger because the luminosity of the sources is weakly limited by the correction. The two Poisson terms are, on the other hand, very similar, since in the mass-dependent model the most massive (and bright) objects that build up the Poisson power are also weakly affected by the correction. The green and orange lines are the two models with redshift dependent escape fractions: the correlation term is strongly reduced in the case of $f_{\mathrm{esc}}(z=4)=0.97$, while the Poisson term, which is generated mainly by low redshift sources, is less affected.

\begin{figure*}
\centering
  \includegraphics[width=5.9cm]{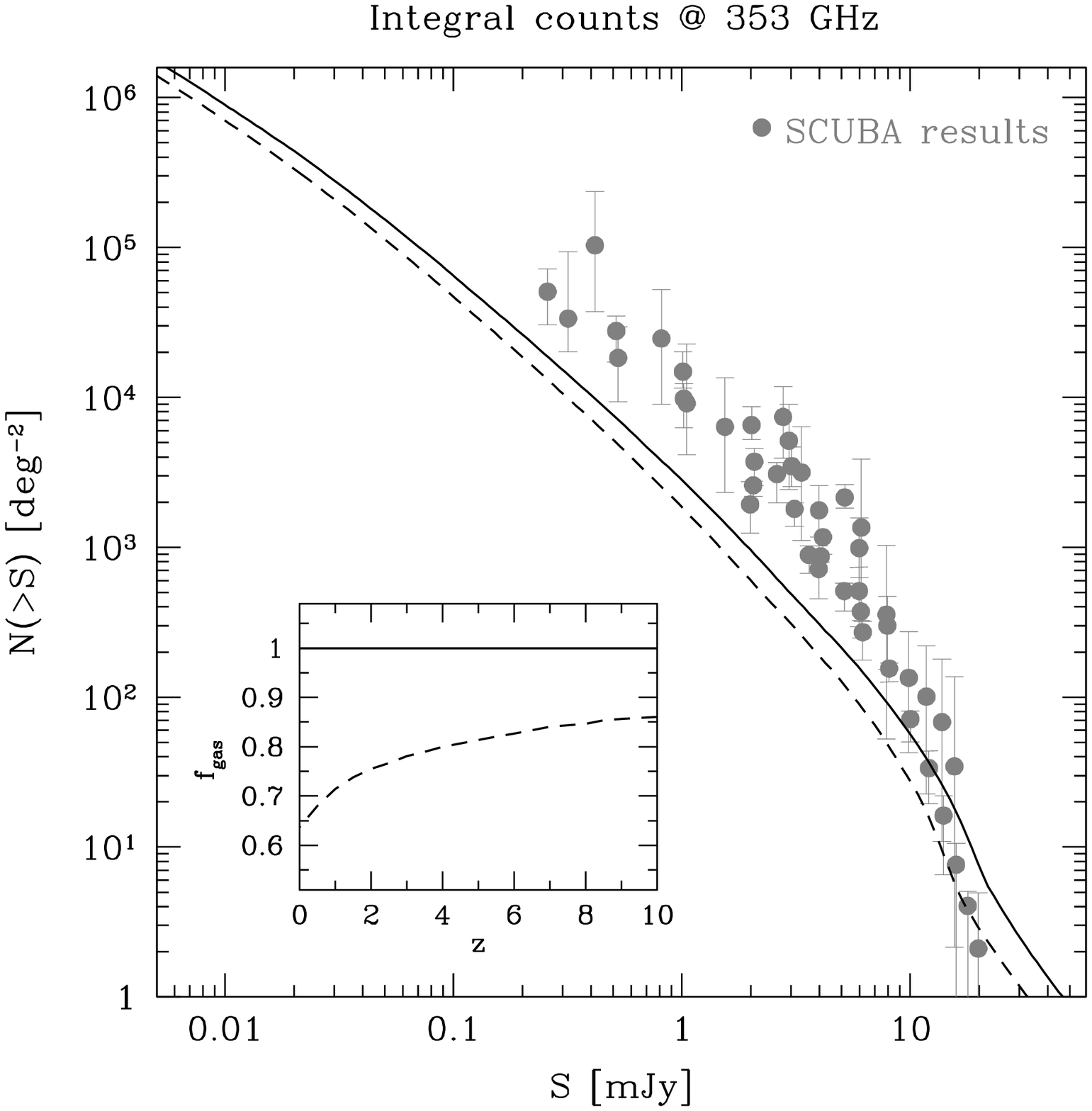}
  \includegraphics[width=5.9cm]{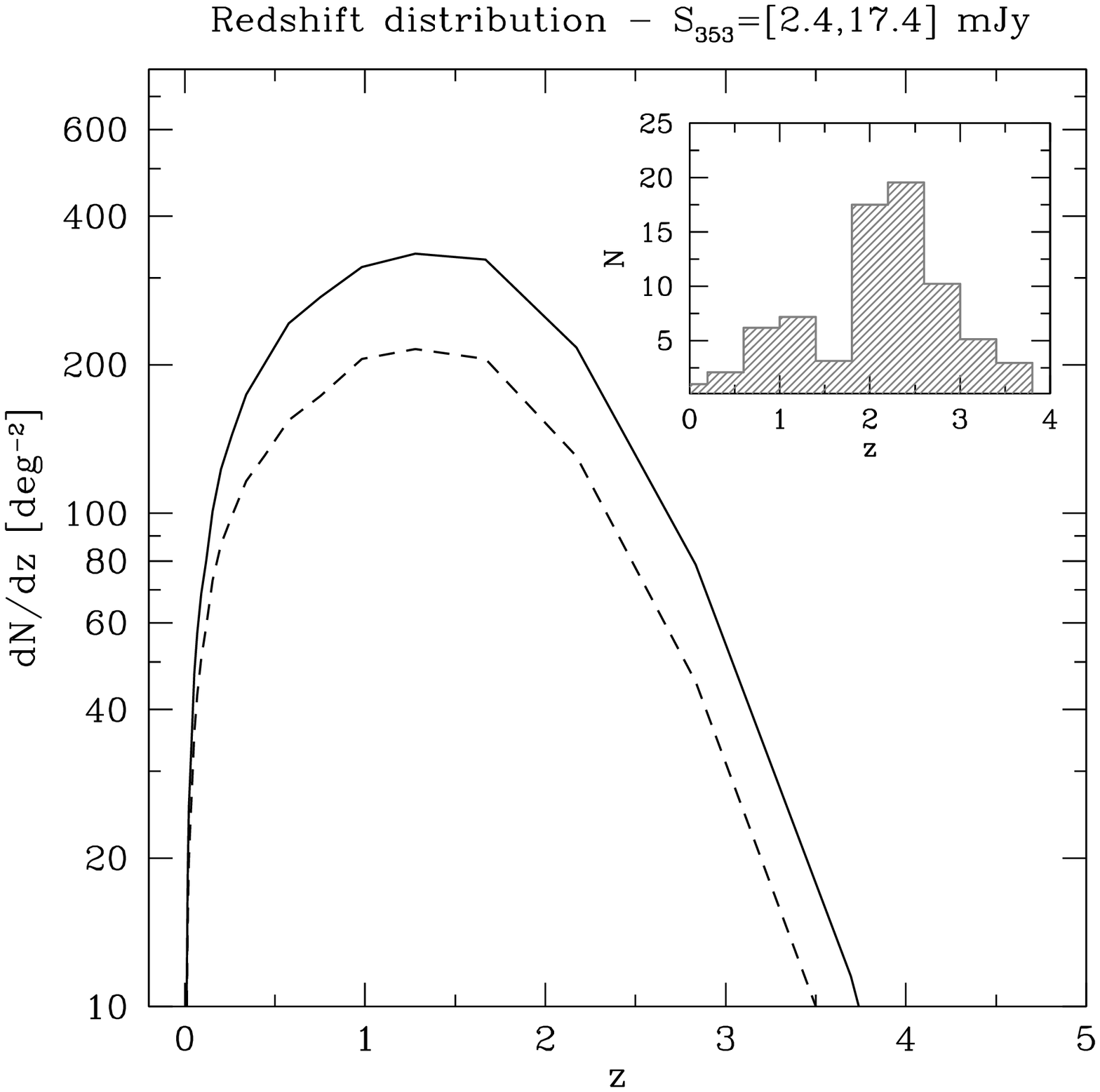}
  \includegraphics[width=5.9cm]{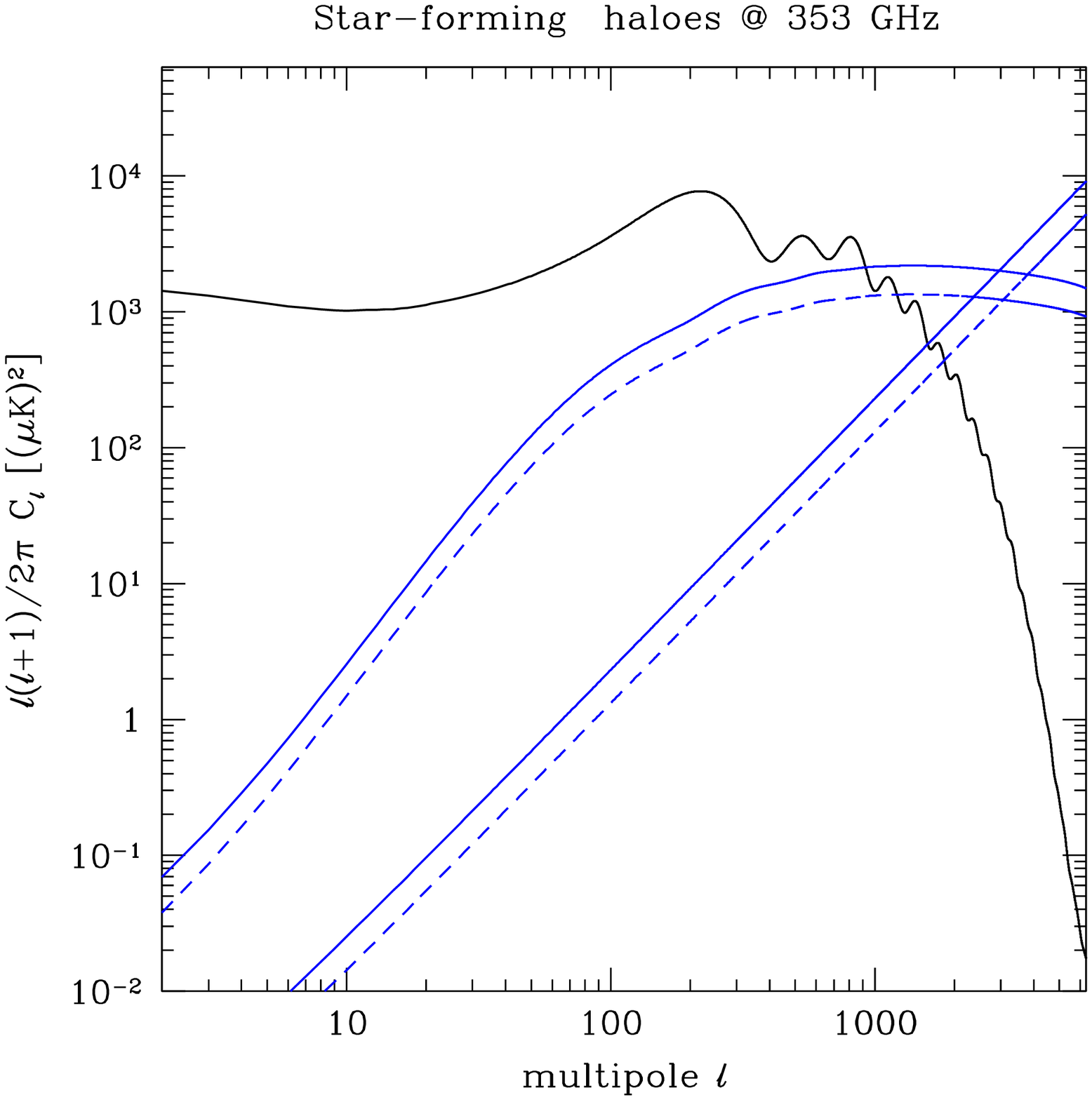}
  \caption{The dependence of our results on the gas fraction. Solid lines refer to the model predictions with $f_{\mathrm{gas}}=1$, dashed lines correspond to the correction in Equation~(\ref{gasfrac}), displayed in the insert in the left panel.}
  \label{fig:fgas}
\end{figure*}

\subsection{Gas fraction}
An additional parameter of the model is the gas fraction inside haloes at different redshift. This quantity establishes the amount of gas that is available at each epoch for the formation of new stars. As explained in Section~\ref{subsec:sfr}, so far we have assumed that all the baryons inside the haloes are in gas form. This approximation is good at high redshift, when the number of stars is small, but it breaks down as more and more baryons are locked into stars.  A simple way to describe this effect is to write the gas fraction at each redshift as
\begin{equation}\label{gasfrac}
f_{\mathrm{gas}}(z)=1-\frac{\rho_{\star}(z)}{\rho_{\mathrm{b}}(z)}=1-\frac{\int_0^{t(z)}\dot\rho_{\star}dt}{\frac{\Omega_{\mathrm{b}}}{\Omega_{\mathrm{m}}}\int M\frac{dn}{dM}dM},
\end{equation}
where $\rho_{\mathrm{b}}$ is the density of cold baryons inside haloes (the integral in the denominator is limited to those masses which are able to cool) and $\rho_{\star}(z)$ is the stellar density, obtained integrating Equation~(\ref{dotrho}) in time. In this way we decrease the contribution of massive objects and reduce the level of the angular fluctuations about $30\%$ (Figure~\ref{fig:fgas}).\\

\section{Conclusions}\label{sec:concl}
We have presented a simple model to determine the space density distribution of haloes which are bright in the sub-millimeter wavelengths due to intense star formation activity, resulting from mergers. We have used the \citet{lc93} approach to compute the merger rate and introduced a characteristic time of star formation for each merger episode, derived from numerical simulations of merging gas-rich disk galaxies. Using the Kennicutt relation we were able to derive the infrared luminosity of each object from its star formation rate.

We assumed a dust emission model based on recent observations of bright infrared galaxies by SCUBA \citep{blainrev}, depending on two free parameters: the emissivity index $\beta$ and the dust equilibrium temperature. We correct the spectral energy distribution taking into account the presence of the CMB thermal bath which becomes non-negligible for high-redshift objects.

With these ingredients, we built estimates for several observables and we calibrate the model in order to obtain a good agreement with available data: the Madau plot for cosmic star formation history and the source counts measured by SCUBA. Furthermore, we are within the limits imposed by FIRAS observations of the Cosmic Infrared Background.

In spite of these promising results, our model can be improved in several aspects. In particular, the merger rate approach is known not to be accurate when compared to $N$-body simulations. Our results depend crucially on several parameters (the escape fraction of ionizing photons, the dust temperature and emissivity index) which might be constrained very precisely through observations of the angular fluctations, at different frequencies and angular scales, by future observations. We are confirming that the dust emission in merging star-forming galaxies is a powerful foreground for the high frequency channels of PLANCK, ACT and SPT.

\begin{acknowledgements}
We thank the anonymous referee for useful remarks.
The authors are grateful to Nicolas Bouch\'e, Rychard Bouwens, Robert Kennicutt, Eiichiro Komatsu, Piero Madau, Ian Smail, Rodger Thompson, Licia Verde and Matias Zaldarriaga for helpful discussions, and to James Bolton for a careful reading of the manuscript. Computations of the power spectra have been performed using the CMBFAST code.
\end{acknowledgements}

\bibliographystyle{aa}
\bibliography{referen.bib}

\appendix

\section{The angular correlation of dust emission generated in star forming haloes}\label{app:powerspec}
In this section we will follow the approach given by \cite{chm06}.

\subsection{Model of measured intensity fluctuations}
When considering sources that are much smaller than the PSF of the observing experiment, it is convenient to explicitly introduce the finite size of the beam and the frequency response of the detector(s). In what follows, our observing experiment will be characterized by a Point Spread Function (PSF) $B(\vec{n})$ ($\vec{n}$ denoting a direction on the
sky) and a frequency response function $\phi (\nu' ,\nu )$, with $\nu$ the central (observing) frequency. For simplicity, we shall assume that both functions have a Gaussian shape, so that 
\begin{equation}
B(\vec{n})=\exp\left[-\frac{\theta^2}{2\sigma_{\mathrm{b}}^2}\right],
\end{equation}
\begin{equation}
\phi(\nu',\nu)=\exp\left[-\frac{(\nu'-\nu)^2}{2\sigma_{\nu}^2}\right].
\end{equation}
The integral over the PSF and the spectral response will be regarded as a normalization factor ${\mathcal N}$:
\begin{eqnarray}
{\mathcal N}&\equiv&\int d\vec{n}\,B(\vec{n})\, d\nu'\,\phi(\nu',\nu)=\nonumber\\
&=&(\sqrt{2\pi}\sigma_{\mathrm{b}})^2\sqrt{2\pi} \sigma_{\nu}\equiv\Omega_{\mathrm{beam}}\,\Delta \nu .
\end{eqnarray}
The brightness intensity fluctuations introduced by the dust emission in haloes can be then written as
\begin{eqnarray}
\Delta I_{\nu}&=&\displaystyle\frac{1}{{\mathcal N}}\displaystyle\int_{0}^{r_{\mathrm{LSS}}} dr\, d\vec{n}\, B(\vec{n})\, d\nu'\,\phi(\nu',\nu)\,\nonumber\\
&&\times\,\displaystyle\sum_j j_{\nu(1+z[r])}\,W(\vec{x}_j-\vec{r})\,a^3(r),
\end{eqnarray}
where $r$ denotes the physical distance from the observer and the index $j$ sweeps over all star forming haloes in the universe, $\vec{x}_j$ being their positions; $j_{\nu (1+z[r])}$ denotes the emissivity at the center of the halo placed at $\vec{x}_j$, and $W(\vec{x}_j - \vec{r})$ is the dust profile and accounts for the dilution of the emissivity as the distance to the halo center increases. Note that due to expansion-induced redshift, the experiment will be sensitive to the emissivity at a frequency $\nu (1+z[r])$. Finally, $a(r)\equiv1/(1+z[r])$ is the scale factor at the epoch given by $r$, and the $a^3$ factor accounts
for the dilution of the intensity as the universe expands. We can replace the sum over $j$ by a volume integral over the star-forming halo density $n(\vec{x})$:
\begin{eqnarray}
\Delta I_{\nu}&=&\displaystyle\frac{1}{{\mathcal N}}\displaystyle\int_{0}^{r_{\mathrm{LSS}}} dr\,d\vec{n}\, B(\vec{n})\,d\nu'\,\phi(\nu',\nu)\nonumber\\
&&\times\,\displaystyle\int d\vec{x}\,n(\vec{x} )\,j_{\nu (1+z[r])}\,W(\vec{x}-\vec{r})\,a^3(r).
\end{eqnarray}
Next we integrate in $\vec{n}$ and $\nu$, in order to obtain the contribution of all
sources within the volume probed by the PSF and the frequency response of the instrument. The vector sweeping this volume will be $\vec{y}$: we use the relation $d\nu'=(\nu/r)\,dy$, and the spectral response $\phi$ becomes a function of $y$, with the maximum at $y = r$.\\ With this in mind, the argument of the function $W$ is now $\vec{x} - \vec{y}$, resulting in
\begin{eqnarray}
\Delta I_{\nu} & = &\frac{1}{{\mathcal N}}\int_{0}^{r_{\mathrm{LSS}}} dr \int d\vec{x}\, n(\vec{x} )\,\left(\frac{\nu}{r^3}\right) \, r^2\,d\vec{n}\, dy \,\phi(y,r)\,j_{\nu (1+z[r])}\nonumber\\
&&\times\,W(\vec{x} - \vec{y})\, a^3(r) =\nonumber \\
\nonumber \\
& = & \frac{1}{{\mathcal N}} \int_{0}^{r_{\mathrm{LSS}}} dr \,\left(\frac{\nu}{r^3} \right)\frac{L_{\nu (1+z[r])}}{4\pi}\,a^3(r)\nonumber\\
&&\times\,\int d\vec{x}\, n(\vec{x} )V(\vec{x} - \vec{r}) .
\end{eqnarray}
In this case, the window function $V(\vec{x}-\vec{r})$ corresponds to the cosmological volume sampled by the finite size of the PSF and the frequency response. The integral on $\vec{x}$ corresponds to a convolution that allows us to rewrite the last equation as
\begin{eqnarray}\label{deltai}
\Delta I_{\nu} &=& \frac{1}{{\mathcal N}} \int_{0}^{r_{\mathrm{LSS}}} dr \,\left(\frac{\nu}{r^3}\right)\frac{L_{\nu(1+z[r])}}{4\pi}\,a^3(r)\nonumber\\
&&\times\,\int\frac{d\vec{k}}{(2\pi)^3}\,n_{\vec{k}} V_{\vec{k}}\,e^{-i\vec{k}\vec{r}}.
\end{eqnarray}
The Fourier modes of the halo density and the window $V$ are given by $n_{\vec{k}}$ and $V_{\vec{k}}$, respectively. Using our model for the PSF and the frequency response, the latter can be written, for each $r$, as 
\begin{equation}
V_{\vec{k}}=\left(\sqrt{2\pi}\sigma_{\mathrm{b}} r \right)^2 \sqrt{2\pi}\, (r/\nu) \,\sigma_{\nu}\,e^{-\frac{k^2\, r^2\sigma_{\mathrm{b}}^2}{2}}e^{-\frac{k^2 (\sigma_{\nu}r/\nu)^2}{2}}.
\end{equation}
The Fourier modes of the halo density will depend on the halo
clustering properties, and will be characterised below. We can introduce the expression for $V_{\vec{k}}$ into
Equation~(\ref{deltai}), and this will remove the dependence on the
normalization factor ${\mathcal N}$
\begin{eqnarray}
\Delta I_{\nu}& = & \int_{0}^{r_{\mathrm{LSS}}} dr \, \frac{L_{\nu (1+z[r])}}{4\pi}\,a^3(r)\nonumber\\
 &&\times\,\int \frac{d\vec{k}}{(2\pi)^3}\, n_{\vec{k}} \,
e^{-\frac{k^2\, r^2\sigma_{\mathrm{b}}^2}{2}}\,e^{-\frac{k^2 (\sigma_{\nu} r/\nu)^2}{2}} \, e^{-i\vec{k}\vec{r}}.
\end{eqnarray}
Further, if we are interested in scales larger than the beam size, the latter expression can be further simplified to
\begin{equation}\label{deltaifinal}
\Delta I_{\nu} \simeq \int_{0}^{r_{\mathrm{LSS}}} dr \, \frac{L_{\nu (1+z[r])}}{4\pi}\,a^3(r)\, n(\vec{r}),
\end{equation}
which is simply an integral of the halo luminosity weighted by the halo density along the line of sight.  From this expression, it is easy to see that the correlation properties of $\Delta I_{\nu}$ will depend on the correlation properties of the star forming haloes.\\
There are, however, two more aspects that need to be taken into account, namely the peculiar velocities of the star forming regions and the gravitational redshift imprinted on the emitted photons on the way to the observer. The former can be observed by simply imtroducing a Doppler term (containing the peculiar velocity of the sources) in the integrand of Equation~(\ref{deltaifinal}), the latter by explicitely writing the time-varying potential fluctuations in such integral.\\
Thus, we rewrite Equation~(\ref{deltaifinal}) as
\begin{eqnarray}
\Delta I_{\nu}&=& \int_{0}^{r_{\mathrm{LSS}}} dr\,\left[\left(1+\frac{\vec{v}\cdot\vec{n}}{c}\right)\frac{L_{\nu(1+z[r])}}{4\pi}\,a^3(r)\,n(\vec{r})\right.\nonumber \\
&&+\,2\left.\left(\int_r^{r_{\mathrm{LSS}}}dr'\,\left(1+\frac{\vec{v}'\cdot\vec{n}}{c}\right)\frac{L_{\nu(1+z[r'])}}{4\pi}\,a^3(r')\,n(\vec{r}')\right)\frac{\dot\phi(r)}{c}\right]\nonumber .\\
\end{eqnarray}
In this new expression, the term $\dot\phi/c$ accounts for the time
derivative of the potential, whereas the dot product
$\vec{v}\cdot\vec{n}$ provides the component of the source peculiar
velocity along the line of sight. If we now argue that we are
interested in scales where perturbations are still in the linear
regime, then we may retain only the first order of the potential,
velocity and density fluctuations that give rise to $\Delta I_{\nu}$:
\begin{eqnarray}
\Delta I_{\nu} &=& \mbox{const}\nonumber\\
&&+\int_{0}^{r_{\mathrm{LSS}}} dr \left[\frac{\vec{v}\cdot\vec{n}}{c}\,\bar{n}(r)\frac{L_{\nu(1+z[r])}}{4\pi}\,a^3(r)+2\frac{\dot\phi}{c}\,\bar{I}_{\nu}^{\mathrm{dust}}(r)\right]\nonumber \\
&&+\int_{0}^{r_{\mathrm{LSS}}} dr \,\delta n(\vec{r})\frac{L_{\nu(1+z[r])}}{4\pi}\,a^3(r)+\mathcal{O}(\delta^2),
\end{eqnarray}
where the average dust intensity $\bar{I}_{\nu}^{\mathrm{dust}}(r)$ is defined as
\begin{equation}
\bar{I}_{\nu}^{\mathrm{dust}}(r)=\int_r^{r_{\mathrm{LSS}}}dr'\, L_{\nu(1+z[r'])}\,a^3(r')\,\bar{n}(r').
\end{equation}
This equation shows that, to the first order, source peculiar
velocities and potential fluctuations will couple to the average
source number density $\bar{n}(r)$, and the correlation properties of
the measured intensities will be ruled by the correlation functions of
each of the three quantities: gravitational potentials, peculiar
velocities and source number overdensities $\delta n(\vec{r})$. We
shall find, however, that the latter will be dominant over the first
two.

\subsection{Correlation of the measured intensity fluctuations}
We first rewrite the change in brightness intensity in terms of an integral of quantities expressed with respect to comoving coordinates
\begin{eqnarray}
\Delta I_{\nu} = \mbox{const}&+&\int_{0}^{r_{\mathrm{LSS}}} dr \left[\frac{\vec{v}\cdot\vec{n}}{c}\,a(r)\,\bar{n}(r)\frac{L_{\nu(1+z[r])}}{4\pi}+2\,\frac{\dot\phi}{c}\,\bar{I}_{\nu}^{\mathrm{dust}}(r)\right]\nonumber \\
&+&\int_{0}^{r_{\mathrm{LSS}}} dr \,a(r)\,\delta n(\vec{r})\frac{L_{\nu(1+z[r])}}{4\pi}a^3(r)+\mathcal{O}(\delta^2)\nonumber .\\
\end{eqnarray}
The comoving number density absorbs the $a(r)^3$ factor, and a new $a(r)$ factor appears from the line element. Note that peculiar velocities are \emph{proper} velocities. Next, we express the star-forming halo number density as an integral over a luminosity distribution function
\begin{equation}
\bar{n}(\vec{x})=\int dL_{\nu(1+z[x])}\frac{dn}{dL_{\nu(1+z[x])}}.
\end{equation}
Finally, we decompose $dn/dL_{\nu(1+z[x])}$ as an integral over the haloes of different masses contributing to the same luminosity interval
\begin{equation}
\frac{dn}{dL_{\nu(1+z[x])}}=\int dM\,\frac{dn}{dM}\,G(M,L_{\nu(1+z[x])}).
\end{equation}
In this equation, the function $G(M,L_{\nu(1+z[x])})$ provides the fraction of haloes present in the mass interval $[M,M+dM]$ (given by the Press-Schechter mass function $dn/dM$) that have recently experienced a major merger, and hence have given rise to significant star forming activity (see Section~\ref{sec:halo}). We then write the spatial correlation function of haloes in the mass range $[M,M+dM]$, $n_{\mathrm{h}}(M,\vec{x})$,
\[
\langle n_{\mathrm{h}}(M_1,\vec{x}_1)\,n_{\mathrm{h}}(M_2,\vec{x}_2)\rangle=\frac{dn}{dM}(M_1,\vec{x}_1)\,\frac{dn}{dM}(M_2,\vec{x}_2)
\]
\[
\phantom{xxxxxxx}+\,\frac{dn}{dM}(M_1,\vec{x}_1)\,\delta_D^3(\vec{x}_1-\vec{x}_2)\,\delta_D(M_1-M_2)
\]
\[
\phantom{xxxxxxx}+\,\frac{dn}{dM}(M_1,\vec{x}_1)\frac{dn}{dM}(M_2,\vec{x}_2)
\]
\begin{equation}
\phantom{xxxxxxx}\times\, b(M_1,z[x_1])\,b(M_2,z[x_2])\,\xi_{\mathrm{m}}(\vec{x}_1-\vec{x}_2).
\end{equation}
The symbol $\delta_D$ stands for Dirac delta and $b(M,z)$ is the mass and redshift dependent bias factor that relates the halo and the matter linear correlation function $\xi_{\mathrm{m}}$ \citep{bias}. The first term on the right hand side accounts for the Poissonian fluctuations in the number counts, whereas the second term accounts for the dependence of the halo number density on the environment.\\
Having this present, one easily finds that the angular correlation function of the intensity fluctuations can be written as
\begin{equation}
\langle\Delta I_{\nu}(\vec{n}_1)\,\Delta I_{\nu}(\vec{n}_2 )\rangle=\sum_l\frac{2l+1}{4\pi}\left(C_l^{\mathrm{P}}+C_l^{\mathrm{C}}\right)P_l(\vec{n}_1 \cdot \vec{n}_2),
\end{equation}
with $P_l(\vec{n}_1 \cdot \vec{n}_2)$ the Legendre polynomia of order $l$ and where $C_l^{\mathrm{P}}$ and $C_l^{\mathrm{C}}$ are the Poissonian and the correlation terms of the $l$-th angular power spectrum multipole. $C_l^{\mathrm{P}}$ is given by
\begin{equation}
C_l^P=\frac{2}{\pi}\int dk\,k^2\left|\int_0^{r_{\mathrm{LSS}}}dr\,a(r)\,j_l(kr)\frac{\sqrt{n\mathcal{L}^2}}{4\pi}\right|^2
\end{equation}
with
\begin{equation}
n\mathcal{L}^2=\int dL_{\nu}\frac{dn}{dL_{\nu}}L_{\nu}^2.
\end{equation}
In practice, this expression is equivalent to
\begin{equation}\label{clp}
\tilde{C}_l^P=\int dS_{\nu}\frac{dN}{dS_{\nu}}S_{\nu}^2
\end{equation}
where $dN/dS_{\nu}$ is the angular number density of sources per spectral flux unit. Due to its simplicity, we preferred to use Equation~(\ref{clp}) in our computations.\\
The correlation term can be expressed in terms of a $k$-space integral of the initial scalar metric power spectrum $P_{\psi}(k)$ times a squared transfer function, as in \cite{cmbfast},
\begin{equation}\label{correl}
C_l^{\mathrm{C}}=\frac{2}{\pi}\int k^2dk\,P_{\psi}(k)\,|\Delta_l (k)|^2,
\end{equation}
with the transfer function $\Delta_l (k)$ given by
\begin{eqnarray}
\Delta_l(k)&=&\int_{0}^{r_{\mathrm{LSS}}} dr\,j_l(kr)\left[S_1(r)\,\delta_k+\frac{dS_2(r)}{dr}\frac{v_k/c}{k}\right.\nonumber\\
&&+\left.S_2(r)\,\frac{\dot{v}_k/c^2}{k}+\frac{\dot\psi+\dot\phi}{c}\,\bar{I}_{\nu}^{\mathrm{dust}}\right].
\end{eqnarray}
In this equation, $j_l(x)$ is the spherical Bessel function of order $l$, $\delta_k$ is the $k$-mode of the dark matter density contrast, $v_k$ is the $k$-Fourier mode of the peculiar velocity field of the emitting sources and $\psi$ and $\phi$ are the two scalar potentials perturbing the FRW metric (in the conformal Newtonian gauge). The functions $S_1(r)$ and $S_2(r)$ are defined by
\begin{eqnarray}
S_1(r)&\equiv&\int dL_{\nu(1+z[r])}\,dM\,G(M,L_{\nu(1+z[r])})\nonumber\\
&&\times\,\frac{dn}{dM}\,\frac{a(r)L_{\nu(1+z[r])}}{4\pi}\,b(M,z[r])
\end{eqnarray}
and
\begin{eqnarray}
S_2(r)&\equiv&\int dL_{\nu(1+z[r])}\,dM\, G(M,L_{\nu(1+z[r])})\nonumber\\
&&\times\,\frac{dn}{dM}\,\frac{a(r)L_{\nu(1+z[r])}}{4\pi},
\end{eqnarray}
respectively.

\end{document}